\documentclass[lettersize,journal]{IEEEtran}
\usepackage{amsmath,amsfonts}
\usepackage{algorithm}
\usepackage{array}
\usepackage[caption=false,font=normalsize,labelfont=sf,textfont=sf]{subfig}
\usepackage{textcomp}
\usepackage{stfloats}
\usepackage{multirow}
\usepackage{url}
\usepackage{verbatim}
\usepackage{graphicx}
\usepackage{cite}
\usepackage{booktabs}
\usepackage{mdframed}
\usepackage{xcolor}

\usepackage{fancyvrb}
\usepackage[noend]{algpseudocode}

\usepackage{xspace}

\definecolor{myCustomGreen}{HTML}{4c97ae}

\newcounter{finding}
\newcommand{\finding}[1]{\refstepcounter{finding}
  \vspace{2mm}
 \begin{mdframed}[linecolor=gray,roundcorner=12pt,backgroundcolor=gray!15,linewidth=3pt,innerleftmargin=2pt, leftmargin=0cm,rightmargin=0cm,topline=false,bottomline=false,rightline = false]
  \textbf{Finding \arabic{finding}:} #1
 \end{mdframed}
 \vspace{2mm}
}

\begin{document}

\title{LLMs are Bug Replicators: An Empirical Study on LLMs' Capability in Completing Bug-prone Code}



\author{Liwei Guo\thanks{Liwei Guo is with Beijing University of Chemical Technology, Beijing, China. Email: liwei.glw@outlook.com.},
        Sixiang Ye\thanks{Sixiang Ye is with Beijing University of Chemical Technology, Beijing, China. Email: yesx.sxye@gmail.com.},
        Zeyu Sun\thanks{Zeyu Sun is with National Key Laboratory of Space Integrated Information System, Institute of Software, Chinese Academy of Sciences, Beijing, China. Email: zeyu.zys@gmail.com.},
        Xiang Chen\thanks{Xiang Chen is with Nantong University, Nantong, China. Email: xchencs@ntu.edu.cn.},
        Yuxia Zhang\thanks{Yuxia Zhang is with Beijing Institute of Technology, Beijing, China. Email: yuxiazh@bit.edu.cn.},
        Bo Wang\thanks{Bo Wang is with Beijing Jiaotong University, Beijing, China. Email: wangbo\_cs@bjtu.edu.cn.},
        Jie M. Zhang\thanks{Jie M. Zhang is with King's College London, London, UK. Email: jie.zhang@kcl.ac.uk.},
        Zheng Li\thanks{Zheng Li is with Beijing University of Chemical Technology, Beijing, China. Email: lizheng@mail.buct.edu.cn.} and
        Yong Liu\thanks{Yong Liu is with Beijing University of Chemical Technology, Beijing, China. Email: lyong@mail.buct.edu.cn.}

        }

\markboth{IEEE TRANSACTIONS ON SOFTWARE ENGINEERING,~Vol.~14, No.~8, August~2021}%
{Shell \MakeLowercase{\textit{et al.}}: A Sample Article Using IEEEtran.cls for IEEE Journals}

\IEEEpubid{0000--0000/00\$00.00~\copyright~2021 IEEE}

\maketitle

\begin{abstract}
Large Language Models (LLMs) have demonstrated remarkable performance in code completion. However, the training data used to develop these models often contain a significant amount of buggy code. Yet, it remains unclear to what extent these buggy instances influence LLMs' performance when tackling bug-prone code completion tasks. To fill this gap, this paper presents the first empirical study evaluating the performance of LLMs in completing bug-prone code. Through extensive experiments on 7 LLMs and the Defects4J dataset, we analyze LLMs' accuracy, robustness, and limitations in this challenging context. Our experimental results show that completing bug-prone code is significantly more challenging for LLMs than completing normal code. Notably, in bug-prone tasks, the likelihood of LLMs generating correct code is nearly the same as generating buggy code, and it is substantially lower than in normal code completion tasks (e.g., 12.27\% vs. 29.85\% for GPT-4). To our surprise, 44.44\% of the bugs LLMs make are completely identical to the pre-fix version, indicating that LLMs have been seriously biased by historical bugs when completing code. Additionally, we investigate the effectiveness of existing post-processing techniques and find that while they can improve consistency, they do not significantly reduce error rates in bug-prone code scenarios. Our research highlights the limitations of current LLMs in handling bug-prone code and underscores the need for improved models and post-processing strategies to enhance code completion accuracy in real-world development environments. 
\end{abstract}


\section{Introduction}

\IEEEPARstart{C}{ode} completion is a pivotal feature of integrated development environments (IDEs) that enhances coding efficiency and accuracy. It automatically completes parts of the code based on a given snippet, functioning either token by token~\cite{10.1145/3510003.3510172} or line by line~\cite{liu2022nonautoregressive}. Code completion plays a crucial role in development, especially with increasingly complex systems and the growing demand for faster development cycles. By accelerating coding efficiency, code completion can help to save time and reduce errors, making it an important tool for developers~\cite{li_toward_2021}.

At an early stage, code completion relied on static analysis and a deterministic set of rules to generate suggestions~\cite{10.1145/2594291.2594321}. With the advancement of deep learning, researchers introduce neural networks to code completion~\cite{liu2022nonautoregressive, guo2022unixcoder,liu2020multitask,Karampatsis_2020,Svyatkovskiy_2019}. More recently, code completion has benefited from the advent of large language models (LLMs)~\cite{achiam2023gpt, nijkamp_codegen2_2023, rozire2024code, lozhkov2024starcoder}, which becomes indispensable in modern software development. GitHub Copilot, Microsoft's AI-powered coding assistant based on GPT-4, has serviced one million paying users~\cite{zdnet_copilot}.



However, LLMs still face many challenges in code completion~\cite{fan_large_2023}. 
In particular, the datasets used to train these models often contain a substantial number of buggy code snippets.
Although many of these bugs are eventually fixed, and the corrected versions coexist with the buggy ones as part of the code's revision history, there is a risk that LLMs may learn — or even memorize — the buggy versions rather than the correct ones, especially in bug-prone scenarios. This could adversely affect the accuracy and reliability of LLMs in code completion.

\begin{figure}[t]
\label{Framework}
\centering
\includegraphics[width=0.45\textwidth]{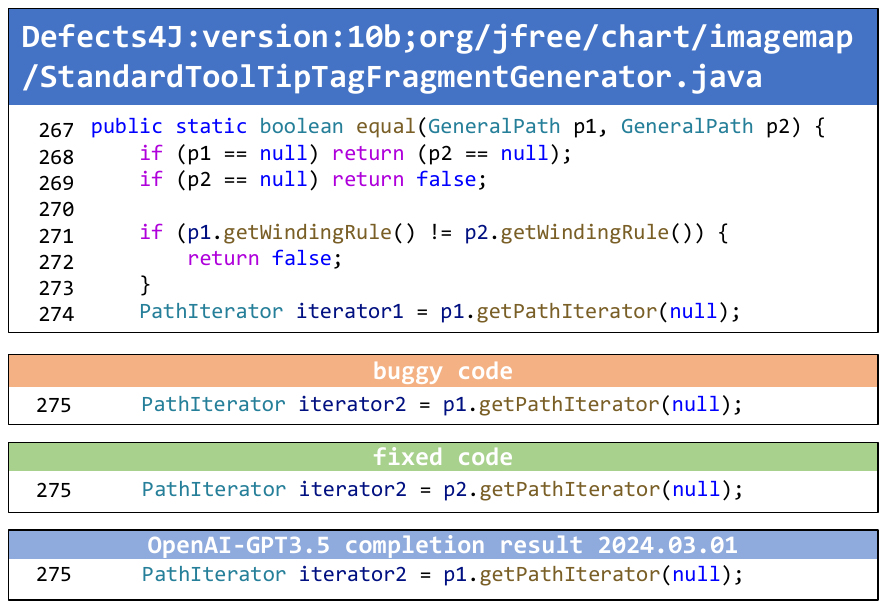}
\caption{Buggy Code Example Generated By OpenAI-GPT3.5}
\label{example1}
\end{figure}

An example of bug-prone code completion is shown in Fig.~\ref{example1}, where OpenAI's GPT-3.5 attempts to complete the code in the \texttt{Generator.java} file. In this example, the original code contains a bug where \texttt{iterator2} is incorrectly assigned using \texttt{p1.getPathIterator(null)} instead of \texttt{p2.getPathIterator(null)}, which should have been used to correctly access the iterator of the \texttt{p2} object. This bug is later fixed by assigning \texttt{iterator2} to \texttt{p2.getPathIterator(null)}, as shown in the ``fixed code'' section. However, when tasked with completing this code, GPT-3.5 still generates the original buggy pattern, incorrectly using \texttt{p1.getPathIterator(null)}.
This example highlights a critical challenge regarding the reliability of LLM-based code completion tools, specifically their ability to complete code that has a history of bugs accurately. It raises important concerns about the extent to which these tools can be relied upon in professional software development environments, where accuracy and dependability are crucial.

\IEEEpubidadjcol

In this study, we aim to investigate the capability of large language models (LLMs) in handling bug-prone code completion tasks. 
These tasks are constructed from code bases with a history of bugs. 
We also investigate the factors that influence LLMs' performance in completing bug-prone code, as well as the effectiveness of post-processing methods in improving such performance.

Specifically, we conduct experiments on 7 LLMs, including ChatGPT (3.5\&4.0\&4o)~\cite{radford2018improving, radford2019language, brown2020language, liu_gpt_2023, ray_chatgpt_2023}, CodeLlama~\cite{touvron2023llama, touvron2023llama2, rozire2024code}, StarCoder~\cite{li2023starcoder, lozhkov2024starcoder},
CodeGEN~\cite{nijkamp2023codegen,nijkamp_codegen2_2023}, 
and Gemma~\cite{gemmateam2024gemma}. To rigorously assess LLM performance in bug-prone code contexts, we construct bug-prone tasks from the Defects4J dataset~\cite{just2014defects4j}, a widely-used benchmark for real-world Java code bugs. 
To further validate our findings, we also conduct additional experiments on the ConDefects dataset~\cite{wu2024condefects}, where we observe consistent patterns, reinforcing the generalizability of our conclusions across different code contexts.

Our experimental results demonstrate that completing bug-prone code poses significantly greater challenges for LLMs compared to standard code completion. Specifically, in bug-prone tasks, LLMs exhibit \textbf{nearly equal probabilities of generating correct and buggy code}, with a substantially lower accuracy than in normal code completion scenarios (e.g., 12.27\% vs. 29.85\% for GPT-4). On average, each model generates approximately 151 correct completions and 149 buggy completions, highlighting the increased difficulty of handling bug-prone contexts.
Moreover, to our surprise, on average, \textbf{44.44\% of the bugs LLMs make are completely identical to the historical bugs}.
For GPT-4o, this number is as high as 82.61\%.
We also observe that the ``if statement'' is the most common construct in bug-prone code. However, completion accuracy is notably low for constructs such as method invocations, return statements, and variable declarations. 
Furthermore, we find that existing post-processing approaches do not fundamentally address the tendency of code completion models to generate incorrect outputs. While post-processing techniques can improve consistency, they do not significantly reduce error rates in the context of bug-prone code.

In summary, our study makes the following contributions:
\begin{itemize}
    \item We provide a systematic evaluation of the ability of seven state-of-the-art LLMs to handle bug-prone code. To the best of our knowledge, this study is the first to comprehensively investigate the effectiveness of LLMs on bug-prone code completion. Our results show that, while LLMs are capable of generating correct code, they exhibit a high likelihood of generating buggy completions in bug-prone code contexts, with a correct-to-buggy completion ratio close to 1:1. This finding highlights a significant limitation of current models in handling complex code dependencies.

    \item We analyze the mechanisms by which LLMs generate errors, with a specific focus on bug-prone code contexts. By examining common code constructs, such as conditionals, loops, and variable declarations, we identify specific weaknesses in the models' handling of certain coding patterns. For instance, we observe that method invocations and return statements are particularly prone to errors, suggesting areas where further training or model refinement may be beneficial.

    \item We evaluate the effectiveness of existing post-processing approaches intended to improve the quality of code completions by LLMs. Our findings indicate that while post-processing can enhance output consistency, it does not significantly reduce the rate of incorrect completions in bug-prone code contexts. This underscores the need for more advanced approaches to address the inherent challenges of bug-prone code completion.
\end{itemize}

To support the open science community, the code and the data used in our study are available at  \url{https://github.com/glwhappen/llms-bug-replicators}. 

\section{Related Work}
\label{sec:relatedwork}
Code completion plays a crucial role in Integrated Development Environments (IDEs)~\cite{bruch2009learning}. It enhances software development efficiency by predicting the remaining code snippets based on previously entered code~\cite{svyatkovskiy2020intellicode, kim2021code, li2024ircoco}. Code completion technology has evolved through several phases.

\subsection{Early Code Completion Techniques}

In the early stages of the research, code completion relied on heuristic rules and static type information~\cite{10.1145/1808920.1808926}, where tools would suggest completions based on the initial characters of variables and function names from a predefined list. This was followed by the adoption of templates and pattern matching to provide more complex code snippet completions.

A significant advancement occurred when researchers recognized the statistical properties of code. Hindle et al.~\cite{6227135} demonstrated that code has predictable statistical patterns, leading to the adoption of statistical language models for code modeling. N-gram models became particularly prevalent~\cite{10.1145/2635868.2635875}, marking a notable evolution in code completion methodologies. However, these methods were less effective for complex code structures and logic, often failing to complete correcct code or taking too much time to generate suggestions.

\subsection{Neural Network-Based Code Completion}

With advancements in neural network technologies, the field of code completion has experienced substantial growth. Modern neural network-based technologies can offer developers contextually relevant and accurate code suggestions. These suggestions consider not just the immediate context~\cite{10.1145/1595696.1595728} and semantics but also the broader patterns and structures within the code.

The scope of code completion ranges from completing the next token or line~\cite{chen2021evaluating} to filling in method and class names~\cite{allamanis_suggesting_2015}, and even extending to the completion of entire programs or projects~\cite{ziegler_productivity_2022}. In development environments, code completion techniques are applied at various levels, with line completion and method completion being the most commonly utilized.

Here, we introduce some classic studies based on neural networks. Svyatkovskiy et al.~\cite{Svyatkovskiy_2019} proposed an LSTM-based code completion system specifically designed to recommend Python method calls. This system is integrated into the Visual Studio Code IDE as part of the IntelliCode extension, enhancing the development environment with advanced method call recommendations.
Liu et al.~\cite{liu2020multitask} developed a Transformer-based neural architecture for code understanding and code generation, utilizing multi-task learning in its pre-training language model. They pre-trained this model with a hybrid objective function that encompasses both code understanding and generation tasks and fine-tuned it to enhance performance in code completion tasks.
Karampatsis et al.~\cite{Karampatsis_2020} introduced a large-scale open-vocabulary neural language model for source code, leveraging the Byte Pair Encoding (BPE) algorithm, beam search, and a caching mechanism. This innovative approach allows for maintaining a low vocabulary size while successfully predicting out-of-vocabulary (OOV) tokens, achieving state-of-the-art performance in token-level code completion.
Guo et al.~\cite{guo2022unixcoder} presented UniXcoder, a unified cross-modal programming language pre-training model that enhances code representation through innovative mechanisms. Utilizing a masked attention matrix with prefix adapters, UniXcoder controls its behavior while leveraging cross-modal contents such as Abstract Syntax Trees (ASTs) and code comments. Additionally, it employs multimodal content and learns code snippet representations via contrastive learning, improving its code completion task capabilities.
Liu et al.~\cite{liu2022nonautoregressive} developed SANAR, a non-autoregressive model with a syntax-aware strategy for line-level completion, which optimizes token pair selection to accelerate and improve completion quality.

\subsection{Large Language Models in Code Completion}

The recent advancements in Large Language Models (LLMs), such as ChatGPT~\cite{radford2018improving, radford2019language, brown2020language, liu_gpt_2023, ray_chatgpt_2023, achiam2023gpt}, CodeGEN~\cite{nijkamp2023codegen, nijkamp_codegen2_2023}, CodeLlama~\cite{touvron2023llama, touvron2023llama2, rozire2024code}, StarCoder~\cite{li2023starcoder, lozhkov2024starcoder}, Gemma~\cite{gemmateam2024gemma}, ChatGLM~\cite{du2022glm, zeng2022glm} Claude~\cite{TheC3}, Grok~\cite{Grok} and DeepSeek-R1~\cite{guo2025deepseek}, have shown great promise in code completion tasks. Trained on vast open-source code repositories~\cite{yin_survey_2023}, they offer developers accurate and intelligent suggestions. For example, after typing a function's initial characters, these models can predict its complete signature~\cite{harte_leveraging_2023}.

However, employing LLMs for code completion introduces also challenges~\cite{ouyang_llm_2023}. The performance of these models largely depends on the quality of their training data, which is primarily sourced from open-source platforms like GitHub~\cite{wei_copiloting_2023}. This data may include buggy code, affecting the models' reliability. Training on such data can lead models to replicate these faults~\cite{fan_large_2023}. A robust model should guarantee high reliability and quality in results. This requires a sophisticated training strategy, including dataset filtering to remove poor coding practices and errors~\cite{nijkamp_codegen2_2023}, and integrating mechanisms for models to identify and avoid such bugs~\cite{white_chatgpt_2023}. Considering these strategies, models can be refined to offer more reliable and best-practice-aligned suggestions~\cite{chen2021evaluating}.

\subsection{Program Repair}

Our research builds upon the same dataset used in the study of program repair~\cite{xia2022practical}. However, we take a distinctly different approach. Program repair involves the identification and fixing of existing code bugs, analyzing buggy code, and applying fixes based on the understanding of known errors and their specific contexts~\cite{goues2019automated, nguyen2013semfix, martinez2016astor, ye2022neural, wei2023copiloting}. In contrast, our study does not equip models with the context surrounding the original bugs. 
Instead, we only provide the bug-prone code snippets and hope that the models recognize these as incomplete code, attempting to complete them.
Our primary focus is to assess the robustness of code completion models when dealing with bug-prone code, rather than their ability to repair known bugs. The key findings of our research indicate that the likelihood of LLMs generating buggy code completions is as high as their probability of generating correct ones.

\section{Study Design}
\label{sec:studydesign}

\subsection{Overview}
This paper aims to answer the following three research questions:
\begin{itemize}
    \item \textbf{RQ1:} \textbf{How do LLMs perform on completing bug-prone code?}
    RQ1 is designed to assess the ability of LLMs to handle bug-prone code completion tasks. We compare the likelihood of LLMs in generating correct code and buggy code. We also compare the performance of LLMs in completing bug-prone code and normal code. 

    \item \textbf{RQ2:} 
    \textbf{What are the characteristics of the tasks where LLMs produce bugs when completing bug-prone code?}
    RQ2 investigates the role of various code constructs (such as conditionals, loops, and variable declarations) in shaping completion accuracy. 
The purpose is to identify areas where LLMs may require further training or refined algorithms to improve their handling of diverse coding patterns.

    \item \textbf{RQ3:} \textbf{Can post-processing methods mitigate the negative effects of bug-prone code on LLM-based code completion?}
    RQ3 evaluates the impact of post-processing approaches on the quality of code completions provided by LLMs, specifically in the context of bug-prone code. Post-processing is crucial for refining raw model outputs. This RQ aims to assess various strategies for post-processing to determine their effectiveness in reducing faults and mitigating the negative effects of bug-prone code in LLM-based code completion.
\end{itemize}

\begin{figure}[ht]
\centering
\includegraphics[width=0.5\textwidth]{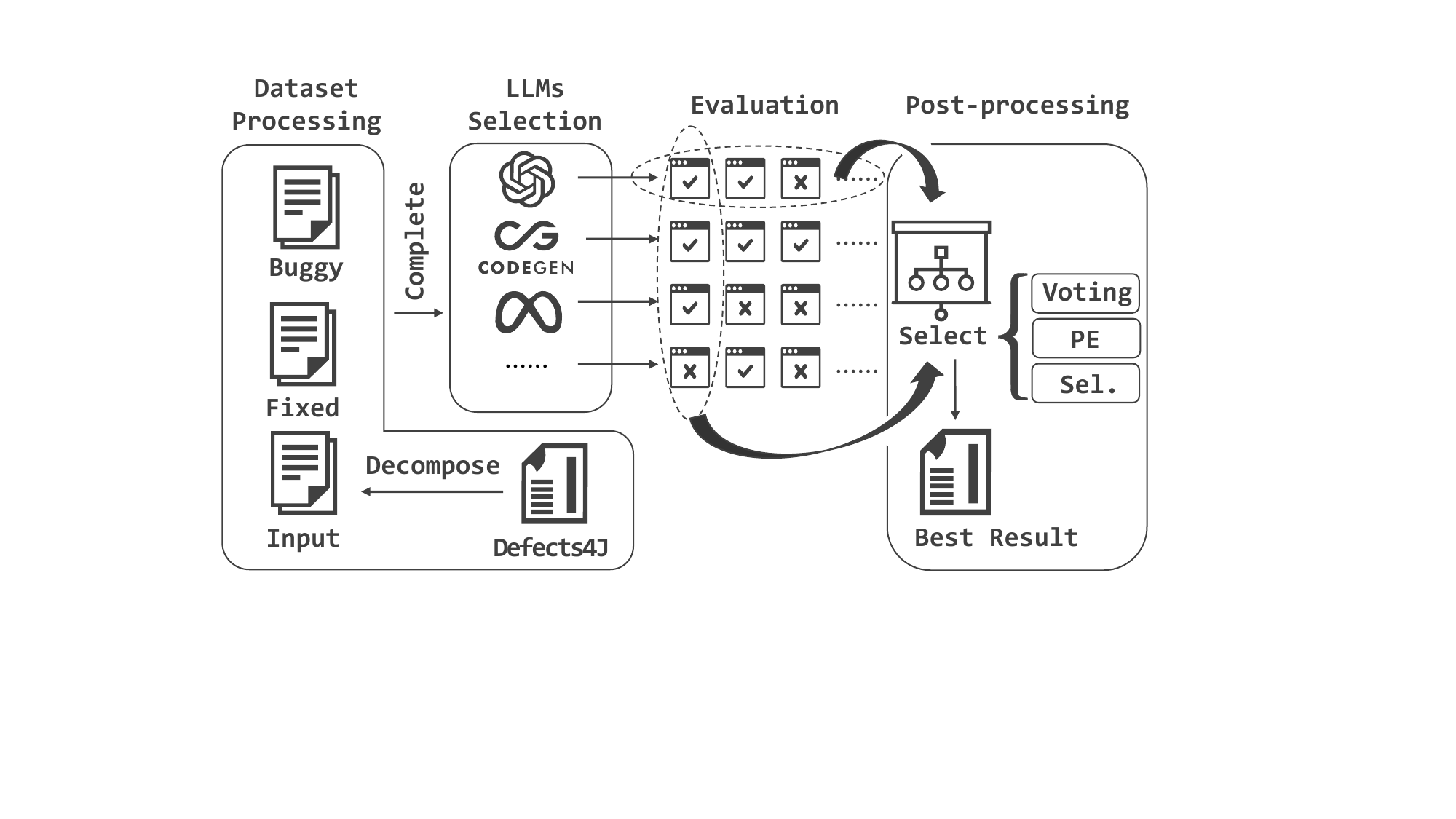}
\caption{Overview of Our Research Methodology}
\label{flow_chart}
\end{figure}

To answer these research questions, we design our methodology, which is shown in Fig.~\ref{flow_chart}.
Specifically, our research methodology consists of four steps. 
(1) \textbf{Dataset Processing}: this step processes the data to facilitate subsequent code completion and evaluation (in Sec.~\ref{section_Dataset}). 
(2) \textbf{LLMs Selection}: this step uses various LLMs to complete the processed input data, generating multiple completion results (in Sec.~\ref{LLMs for Code Completion}). 
(3) \textbf{Evaluation}: this step is performed by evaluating completions of LLMs using our adopted metric (in Sec.~\ref{Evaluation Metrics}). 
(4) \textbf{Post-processing}: this step employs three post-processing approaches to refine completion results (in Sec.~\ref{Post-processing}). 



\subsection{Dataset Processing and Task Construction}
\label{section_Dataset}

Identifying and collecting data for code that is inherently prone to bugs is challenging, as it requires a systematic way to determine and validate its bug-proneness across diverse scenarios. Instead, we leverage existing datasets of real-world software defects, where the bugs are already identified and fixed. This approach is justified because historically buggy code provides strong empirical evidence of patterns and structures that are more susceptible to defects. By analyzing the pre-fix versions of such code, we can approximate the characteristics of bug-prone code, ensuring that our study is grounded in realistic and well-documented software failures. Moreover, using established bug datasets enhances reproducibility and comparability with prior research, allowing for more robust evaluations of how code completion models handle bug-prone scenarios.

We use the Defects4J 2.0 dataset~\cite{just2014defects4j} to construct the main bug-prone code completion tasks, which comprises 697 defects from 16 real-world open-source Java projects on GitHub. This dataset is extensively used for evaluating fault localization~\cite{sarhan_survey_2022, chen2023future, yang2024large, ghanbari2023mutation, wu2023large, 10.1145/3510003.3510073} and program repair~\cite{durieux2019empirical, 7985681, Xiong_2018, Martinez_2016, 8812054, Motwani2018}, 
underlining its reliability and relevance for bug-related research. 

With this dataset, we systematically reconstruct code scenarios with known historical bugs.
We exclude cases where fixes involve adding or deleting code and concentrate on defects that involve modifications to existing lines of code, as this allows us to accurately compare between the completed code and historically buggy code.
Following this criterion, 546 defects are kept for our analysis.

For each bug, we retain the code preceding the buggy line as the context for LLMs to perform code completion. For example, if a defect occurs at line 50, the first 49 lines of code are provided to the LLM, allowing it to generate the remaining code while assessing its susceptibility to historical bugs.



We take the first line LLMs complete for analysis, because in Defects4J, 76.76\% of patches are single-line~\cite{just2014defects4j}. 
Moreover, single-line code completion is the most common in practical coding environments~\cite{wang2023practitioners}.
The completed line is then compared with the corresponding line in both the pre-fix and fixed versions in Defects4J. 
In Section~\ref{sec:discussion}, we investigate whether analyzing just one-line of completed code is a threat to our conclusions.


\subsection{LLMs}
\label{LLMs for Code Completion}

Our next step is to select LLMs for evaluation. We consider both proprietary and open-source LLMs to guarantee the generalization of our empirical findings. 
As a result, we select seven state-of-the-art LLMs. 
For proprietary LLMs, we select OpenAI-GPT4o~\cite{achiam2023gpt}, OpenAI-GPT4~\cite{achiam2023gpt}, OpenAI-GPT3.5~\cite{brown2020language}. For open-source LLMs, we select CodeLlama-7B-hf~\cite{rozire2024code}, Gemma-7B~\cite{gemmateam2024gemma} and StarCoder2-7B~\cite{lozhkov2024starcoder}. We also notice that reasoning models (e.g., DeepSeek R1~\cite{guo2025deepseek}) perform well but require extended reasoning time, making them less suitable for prompt code completion. For these models, the analyses are in Sec.~\ref{sec:reasoning}.

In our study, for ChatGPT, we use the official API provided by OpenAI. As for the Gemma, CodeLlama, StarCoder, and CodeGEN models, we use the API provided by HuggingFace~\cite{huggingface}.
To guide the interactive dialogue model in generating relevant responses, we use the following prompt: ``I need to complete the following code snippet. Please provide only the continuation of the code, without any additional comments or repeating the original code. The code is as follows:\{code\}''.

\subsection{Evaluation}
\label{Evaluation Metrics}

After data processing and model selection, we feed code snippets into LLMs for code completion. Since we are only completing code lines, the resulting code is not executable as a whole. Consequently, execution-based evaluation methods are not applicable here. Instead, we aim to evaluate the generated code against the buggy code and the fixed code.

To ensure a reasonable assessment of code completions, we use an evaluation metric that combines the Longest Common Subsequence (LCS) and Levenshtein Edit Distance (LED). This metric involves a comparative analysis of the completed code against both the correct and buggy code present in the Defects4J dataset~\cite{liu2022nonautoregressive}. We do not use existing metrics like CodeBLEU~\cite{ren2020codebleu} and CodeBERT scores~\cite{zhou2023codebertscore} because they are designed to measure the similarity of multi-line code. In our study, we focus primarily on single-line code, which is often very short. Therefore, these metrics are not suitable for our study (the experimental analysis is in Section~\ref{sec:rq2}). 

This evaluation metric assesses two main scores: the ``bug\_proximity\_score" and the ``fix\_proximity\_score." The ``bug\_proximity\_score" measures the similarity between the completed code and the original buggy code, while the ``fix\_proximity\_score" evaluates the similarity between the completed code and the fixed version.
To compare, each score assesses the similarity between a single line of code $s$ and samples of fixed or buggy code $t$. We combine the widely used LCS and LED. LCS computes the longest subsequence common to both sequences under comparison. We normalize the LCS by dividing its value by the length of the longer string between string $s$ and string $t$. 
LED computes the minimum number of edits (such as insertions, deletions, and substitutions) needed to transform one string into another. We normalize the LED by calculating the inverse proportion of the LED value to the length of the longer string, as a lower edit distance indicates higher similarity. 


Since these two metrics are equally important, we assign the same weight for both. Given a line of completed code $s$,  fixed code $t'$, and  buggy code $t$, the formulas for calculating the similarity of the code completion results are shown as follows:
$\text{bug\_proximity\_score}(s, t)  = \frac{\text{LCS}(s, t) 
+ \text{LED}(s, t)}{2}$ and $\text{fix\_proximity\_score}(s, t') = \frac{\text{LCS}(s, t') + \text{LED}(s, t')}{2}$.

We find that when the score is below $threshold$, the difference between the two texts becomes very significant. Therefore, we set the $threshold$ at 0.4 to distinguish whether the two texts are completely dissimilar in both semantics and tokens based on the experimental preliminary analysis. The evaluation metric classifies the results into the following five categories:

\begin{itemize}
    \item \textbf{Bug-identical}: If $\text{bug\_proximity\_score}(s, t) = 1$, indicating that the generated code $s$ is identical to the buggy code $t$. This implies that the model replicates the historical bug without any correction.
    
    \item \textbf{Correct-identical}: If $\text{fix\_proximity\_score}(s, t') = 1$, indicating that the generated code $s$ exactly matches the fixed code $t'$. This demonstrates that the model produces an error-free version that perfectly corresponds with the historical correct fix.
    
    \item \textbf{Non-compliant}: If both $\text{bug\_proximity\_score}(s, t) < threshold$ and $\text{fix\_proximity\_score}(s, t') < threshold$.  This indicates that the generated code $s$ does not sufficiently align with either the historical bug or the historical correct fix, potentially resulting in irrelevant or nonsensical code snippets.
    
    \item \textbf{Bug-close}: If $\text{bug\_proximity\_score}(s, t) > \text{fix\_proximity\_score}(s, t')$, $\text{fix\_proximity\_score}(s, t') \geq threshold$, and $\text{bug\_proximity\_score}(s, t) \neq 1$. This indicates that the generated code $s$ is closer to the historical bug $t$, indicating that it resembles the historical buggy code with modifications and suggesting that the LLM may tend to replicate historical bugs.
    
    \item \textbf{Correct-close}: If $\text{fix\_proximity\_score}(s, t')$ is greater than $\text{bug\_proximity\_score}(s, t)$, $\text{bug\_proximity\_score}(s, t)$ meets or exceeds threshold, and $\text{fix\_proximity\_score}(s, t') \neq 1$, then the generated code $s$ is deemed Correct-close, meaning it closely aligns with the historical correct fix despite minor discrepancies.
\end{itemize}

In particular, our evaluation metric is designed not only to compare the generated code with both the buggy and fixed versions but also to detect if LLMs tend to replicate historical bugs. The bugs in our dataset, appearing in official open-source releases, are often subtle and challenging for developers to identify. In contrast, more obvious bugs may not offer the same insight into the models’ nuanced behavior.

\subsection{Post-processing Approaches}
\label{Post-processing}


Post-processing methods have been widely adopted to enhance the performance of existing models~\cite{sun2020automatic,li2023cctest,dai2023uncovering,zhang2023coder,yin2019reranking}. 
It uses additional techniques after the initial code generation to refine the result and enhance its accuracy. 
Therefore, in this study, we also explore three widely used post-processing approaches.
In particular, we let LLMs generate multiple completed code snippets as candidates. 
The focus of post-processing is to identify and select the best result among all these candidates.

The post-processing approaches we have explored are: 

1) \textbf{Majority Voting}: This approach involves generating multiple responses from the LLM and uses a voting mechanism to select the result that most closely aligns with the majority.

2) \textbf{Prompt Engineering}: This approach involves generating multiple responses from the LLM and allowing the LLM to select the best result from these generated outputs.

3) \textbf{Candidate Selection Model}: This approach involves generating multiple responses from the LLM and training a custom model to select the best result based on predefined criteria.

These approaches are adjusted and implemented to enhance the effectiveness of the selection process, ensuring that the most accurate and useful code snippet is identified for use. In the rest of this subsection, we introduce these post-processing approaches in more details.

\subsubsection{Majority Voting}


Majority voting employs a voting rule to select the code completion that is most similar to the others from multiple results~\cite{li2023cctest,zhou2024don, lewkowycz2022solving, huang2022large}. The purpose of this approach is to exclude outliers and select the most representative result. For our investigated bug-prone code completion, we generated several code completion results using a LLM, and then calculated the sum of similarities between each code completion and the rest to assess the average similarity of each result. Ultimately, the code completion with the highest average similarity was selected. Below is a detailed explanation of this process:

Suppose we have \(N\) code completion results, each denoted as \(C_i\), where \(i = 1, 2, \ldots, N\). These results are generated by the large model, and we aim to calculate the sum of similarities between each result and all other results. The similarity function can be represented as \(sim(C_i, C_j)\), quantifying the similarity between two code completions \(C_i\) and \(C_j\). The calculation method for \(sim\) is the same as that used for \(bug\_proximity\_score\) and \(fix\_proximity\_score\) in `Evaluation Metrics,' except that it compares the similarity between two code completions.

Formally, for each code completion \(C_i\), its average similarity \(S_i\) with all other completions is calculated via 
$S_i = \frac{1}{N-1} \sum_{\substack{j=1 \\ j \neq i}}^{N} sim(C_i, C_j)$.
Here, \(S_i\) is the average similarity of \(C_i\) with all other results. Finally, we select the code completion \(C_i\) with the highest \(S_i\) as the best code completion result.

\subsubsection{Prompt Engineering}

\begin{figure}[h]
\centering
\includegraphics[width=0.5\textwidth]{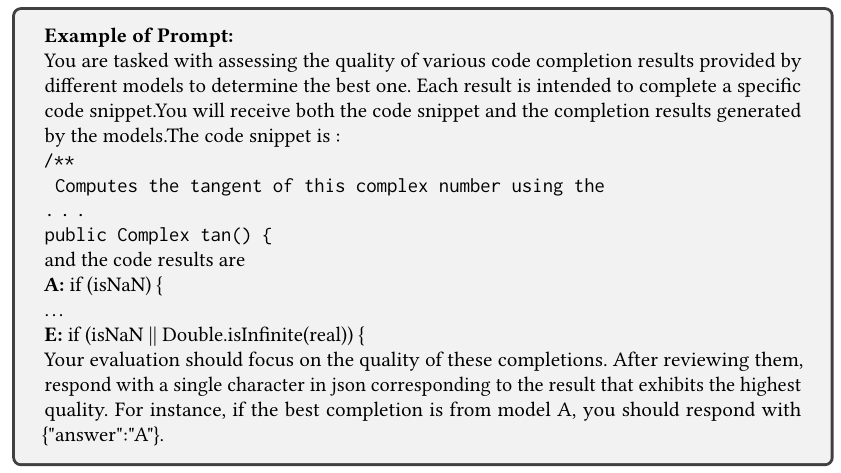}
\caption{Example of Prompt}
\label{prompt}
\end{figure}

Prompt Engineering (PE) has recently gained popularity as an approach to enhancing the abilities of LLMs without altering their model weights, thereby improving their accuracy on specific tasks~\cite{wei2022chain, wang2023prompt}. In our approach, we use LLMs to generate multiple completions and then select the best result from these outputs. Specifically, we employ a particular prompt for selection, which is designed based on the work of Guo et al.~\cite{guo2024sample}, who found that placing instructions at the beginning of the prompt yields the best task performance.
Additionally, specifying the output format as JSON can enhance the stability of the result format, likely due to its structured nature. With this prompt, we input code snippets and their multiple generated code snippets, letting the LLM choose the optimal one. We provide an example of our used prompt as follows. A prompt example is shown in Fig.~\ref{prompt}.

\subsubsection{Candidate Selection Model}


This approach involves generating multiple responses from the LLM and using a custom-trained selection model to choose the best result. Specifically, we develop the selection model through additional training of the CodeT5+ model~\cite{wang2023codet5+}.
The training used a bug-fix dataset from Java projects~\cite{zhu2021syntax}. The dataset contains 103,585 patches, focusing only on patches that modified or introduced a single statement. 


These patches are processed to extract pairs of data representing pre and post modifications, consistent with the segmentation method described in Sec.~\ref{section_Dataset}, regarded as a bug and its corresponding fix. The CodeT5+ model is employed for training to predict whether a line of code is defective. The trained model is subsequently used to predict the likelihood of defect-free code from completed code in large language models, assigning probability scores to evaluate and select the most promising result based on the highest score.

\subsection{Running Platform}

We run the experiments on a server with four NVIDIA A100 Tensor Core GPUs, each with 80 GB of memory.

\section{Results and Analysis}
\label{sec:results}

\subsection{RQ1: How do LLMs perform on completing bug-prone code?}

To answer RQ1, we assess the quality of code completion results from the selected LLMs, focusing on single-line code completions and their evaluation compared to correct or buggy code. 
Each model completes 546 code snippets, which are then categorized based on our evaluation metric.
For all experiments, we repeat five times for each model. 



The results are presented in Table~\ref{completion results}, where each row corresponds to an LLM, and each column displays the results for the corresponding categories, as assessed by our evaluation metric. Each cell in the table shows both the number and the corresponding percentage for each category. Specifically, this table also includes the results for the 'normal' code (in Type), which consists of 546 code snippets selected from the correct code snippets in the same code file as the corresponding bug-prone code snippets. To ensure a fair comparison, the code structures (e.g., lines in if statements or while loops) of each selected snippet match those of the original 546 code snippets\footnote{If structurally equivalent normal code cannot be found within the same file, we expand our search to subsequent files within the dataset.}. For these snippets, only the correct version of the completion exists, and we measure the similarity score of each code completion against the correct implementation (referred to as $\text{fix\_proximity\_score}$).

\begin{table*}[h]
\centering
\caption{Analysis of code quality completed by different LLMs}

\resizebox{0.98\linewidth}{!}{ 
\begin{tabular}{llrrrrrrr}
\toprule
\textbf{LLMs} & \textbf{Task Type} &
\textbf{\begin{tabular}[c]{@{}c@{}}Correct-identical\end{tabular}} & \textbf{\begin{tabular}[c]{@{}c@{}}Bug-identical\end{tabular}} &
\textbf{\begin{tabular}[c]{@{}c@{}}Correct-close\end{tabular}} & \textbf{\begin{tabular}[c]{@{}c@{}}Bug-close\end{tabular}} &  
\textbf{Non-compliant} &  
\textbf{\begin{tabular}[c]{@{}c@{}}Bug-identical ratio\\ (against all bugs)\end{tabular}} \\ \midrule
\multirow{2}{*}{OpenAI-GPT4o} & bug-prone & 151 (27.66\%) & 152 (27.84\%) & 26 (4.76\%) & 32 (5.86\%) & 185 (33.88\%) & 82.61\% \\ 
 & normal & 183 (33.52\%) & - & 189 (34.62\%) & - & 174 (31.87\%) & - \\ \midrule
\multirow{2}{*}{OpenAI-GPT3.5} & bug-prone & 125 (22.89\%) & 91 (16.67\%) & 101 (18.50\%) & 87 (15.93\%) & 142 (25.82\%) & 51.12\% \\
 & normal & 248 (45.42\%) & - & 145 (26.56\%) & - & 153 (28.02\%) & - \\\midrule
\multirow{2}{*}{OpenAI-GPT4} & bug-prone & 67 (12.27\%) & 59 (10.81\%) & 107 (19.60\%) & 95 (17.40\%) & 218 (39.93\%) & 38.31\% \\ 
 & normal & 163 (29.85\%) & - & 170 (31.14\%) & - & 213 (39.01\%) & - \\\midrule
\multirow{2}{*}{CodeLlama-13B-hf} & bug-prone & 57 (10.44\%) & 42 (7.69\%) & 76 (13.91\%) & 105 (19.23\%) & 266 (48.72\%) & 28.57\% \\
 & normal & 142 (26.01\%) & - & 179 (32.78\%) & - & 225 (41.21\%) & - \\\midrule
\multirow{2}{*}{Gemma-7B} & bug-prone & 34 (6.23\%) & 18 (3.30\%) & 88 (16.12\%) & 102 (18.68\%) & 304 (55.68\%) & 15.00\% \\ 
 & normal & 101 (18.50\%) & - & 204 (37.36\%) & - & 241 (44.14\%) & - \\\midrule
\multirow{2}{*}{StarCoder2-15B} & bug-prone & 31 (5.68\%) & 27 (4.95\%) & 64 (11.72\%) & 88 (16.12\%) & 333 (61.00\%) & 23.48\% \\ 
 & normal & 85 (15.57\%) & - & 177 (32.42\%) & - & 284 (52.01\%) & - \\ \midrule
\multirow{2}{*}{CodeGEN-350M} & bug-prone & 30 (5.49\%) & 26 (4.76\%) & 97 (17.77\%) & 119 (21.79\%) & 274 (50.18\%) & 17.93\% \\
 & normal & 91 (16.67\%) & - & 189 (34.62\%) & - & 266 (48.72\%) & - \\

\bottomrule
\end{tabular}
}
\label{completion results}
\end{table*}




We analyze the results from three perspectives: 1) the quality of completions, which evaluates the performance of models in completing; 2) memorization, where models completely remember incorrect answers and repeat them; and 3) the comparison between bug-prone and normal code completion tasks, highlighting how models perform on error-prone versus correct code.

\paragraph{The quality of completions}
To evaluate the quality of completions,
from this table, we can observe that the proportion of correct identical completions is low across all models for bug-prone data. The highest is 27.66\% for OpenAI-GPT4o, while the lowest is 5.49\% for CodeGEN-350M. This indicates a challenge in generating perfect code snippets. However, when combined with correct close completions, the models seem to perform better, though variability is evident. For instance, OpenAI-GPT3.5 achieves a total correct rate (Correct-close + Correct-identical) of 41.39\%, whereas StarCoder2-15B only reaches 17.4\%. 

Such completions show that OpenAI-GPT4o and OpenAI-GPT3.5 perform better compared to other models. However, when we consider the bug completions, a deeper analysis is required. For instance, while OpenAI-GPT4o achieves a relatively high correct completion rate of 32.42\% (combining Correct-close and Correct-identical) ,it also has a significant proportion of buggy completions at 5.86\% for Bug-close and 27.84\% for Bug-identical. Similarly, OpenAI-GPT3.5, despite having the highest correct completions, also has a significant rate of buggy completions at 15.93\% for Bug-close and 16.67\% for Bug-identical.

These results show an issue: even models that perform well in generating correct code are also prone to generating substantially incorrect outputs. This may be due to the models' sensitivity to the noise in the training data, which can include both correct and incorrect coding patterns.

We also find that the ratio of correct completions to buggy completions is close to 1:1 for several models in the bug-prone type (the average ratio is 1.01:1.00). This proximity in ratios indicates that while LLMs have significant potential in generating correct code in the bug-prone context, they are equally prone to generating buggy outputs, reflecting inherent challenges in model training and application. In particular, OpenAI-GPT3.5 stands out with the highest rate of correct completions at 45.42\% in the normal context and 22.89\% in the bug-prone context, but it is closely followed by a significant 32.6\% buggy completions in the bug-prone context.

This indicates a critical need for improved training methods, better quality assurance in training data, and enhanced post-processing approaches to ensure that the benefits of using LLMs for code completion outweigh the risks associated with their current limitations. Addressing these challenges is essential for advancing the reliability and efficacy of LLMs in code completion tasks.

\begin{figure}[t]
\centering
\includegraphics[width=0.45\textwidth]{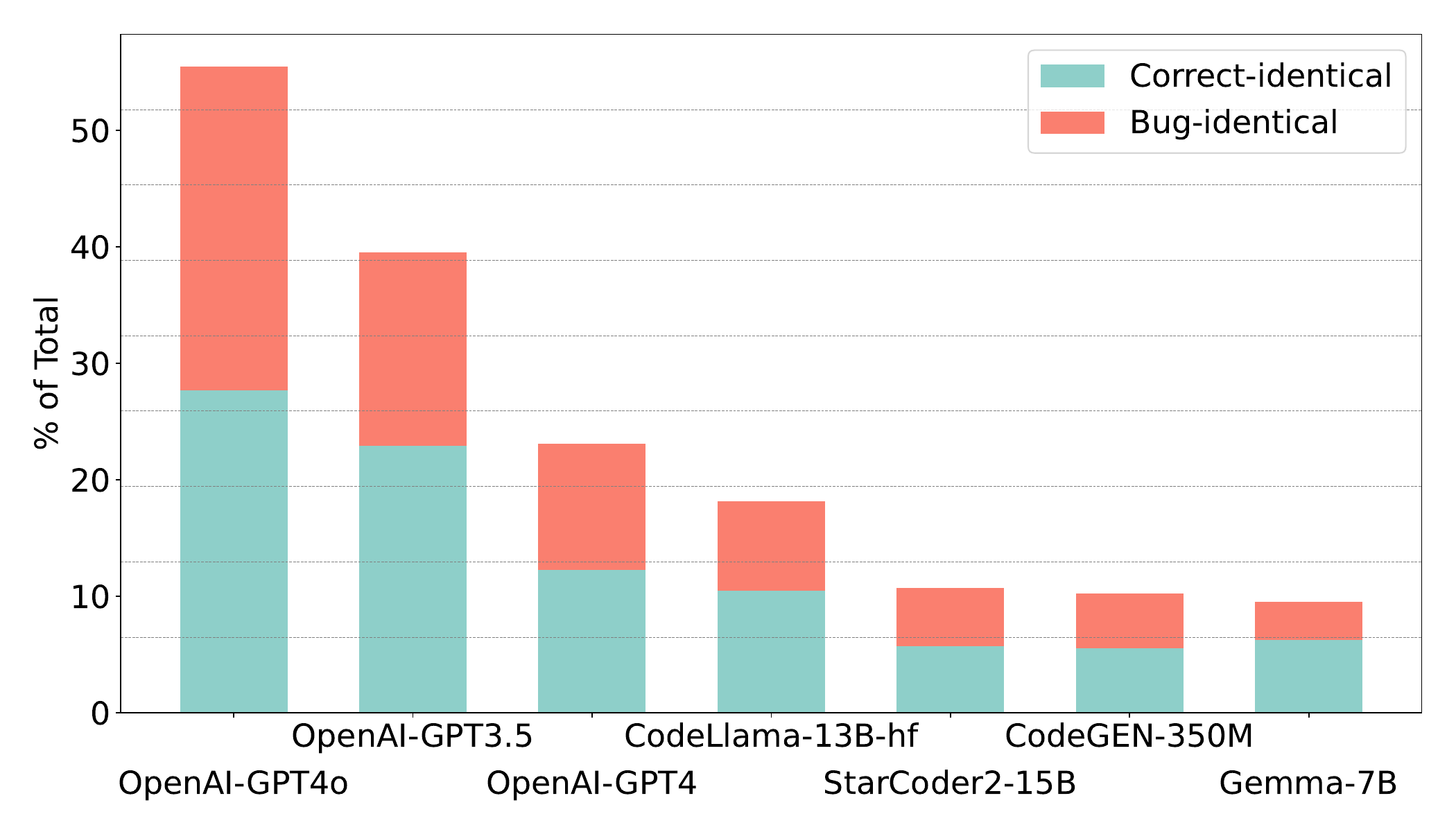}
\caption{Analysis of Code Memorization Across Different LLMs}

\label{memory}
\end{figure}

\paragraph{Memorization}
Further, we change the aspect to the memory of these LLMs. Within our evaluation, completions that align perfectly with either the correct or buggy code snippets provided suggest that a model may have memorized these specific sequences, hinting that its training data likely encompasses projects similar to those in the Defects4J dataset or analogous compilations. This observation implies that when LLMs generate identical results, they are likely regenerating learned patterns from their training datasets rather than applying learned coding principles to new code completion scenarios. To quantify this phenomenon, we analyze the extent to which various models memorize code from the training dataset, with our analyzed results illustrated in the subsequent Fig.~\ref{memory}. 

In this figure, we observe that the memorization rates for the models range between 10\% and 55\%. OpenAI's GPT-4o and GPT3.5 stand out for their high levels of code memorization, with proportions reaching 55.5\% and 39.56\%, respectively. This suggests that the training datasets for both models likely contain code that closely resembles our experimental subjects. In contrast, GPT-4 demonstrates a lower tendency toward code memorization, likely due to its training on a more diverse collection of code. This diversity may allow GPT-4 to approach code completion tasks with a broader set of strategies rather than simply replicating specific results from the training data. The memorization rates for other models fall between 10\% and 20\%, highlighting significant variation in how different models handle code replication.

When we specifically focus on the memorization of buggy code, we find that a significant portion of the bugs LLMs generate are directly copied from their training data. For instance, OpenAI's GPT-4o has a memorization rate of 27.84\%, meaning that nearly 28\% of the bugs it produces are identical to the buggy code in its training set. GPT-3.5 follows with 16.67\%, while Gemma-7B has the lowest memorization rate at 3.30\%. 

Building on the findings, we further investigate how generated bug completions reflect historical buggy patterns. Among all bug completions (Bug-identical and Bug-close), the Bug-identical Ratio quantifies the proportion of generated bug completions that are exactly identical to historical buggy code. The results are listed in Table~\ref{completion results}. We find that the Bug-identical Ratio varies markedly across models, ranging from 15\% to 83\%. For example, OpenAI’s GPT-4o exhibits a ratio of 82.61\%, and GPT-3.5 follows with 51.12\%, implying that a significant portion of their buggy outputs are direct copies of known errors from the training data. In contrast, Gemma-7b’s notably low ratio of 15.00\% suggests that its buggy completions are more often merely token-wise similar to historical bugs rather than exact reproductions. This indicates that models with higher Bug-identical Ratios are more reliant on memorizing and reproducing buggy patterns from their training data, which may hinder their ability to innovate and generate error-free code.

These findings further support our hypothesis that the presence of buggy codes in the training dataset can profoundly affect the quality of code completions generated by LLMs, indicating an urgency to identify and remove noises in the training datasets, which can help to improve code completion quality.

\paragraph{Comparison between bug-prone and normal code completion tasks.}

After examining the challenges of bug-prone code completion, we now focus on comparing bug-prone and normal code completion tasks.

As shown in Table~\ref{completion results}, the accuracy of generating correct completions is significantly higher for normal code compared to bug-prone code. For instance, OpenAI-GPT3.5 achieves a high rate of 45.42\% Correct-identical completions for normal code, while OpenAI-GPT4 achieves a combined Correct-close and Correct-identical rate of 60.99\%. 
This highlights the increased difficulty in completing bug-prone code.

\begin{figure*}[t]
    \centering
    \includegraphics[width=0.7\textwidth]{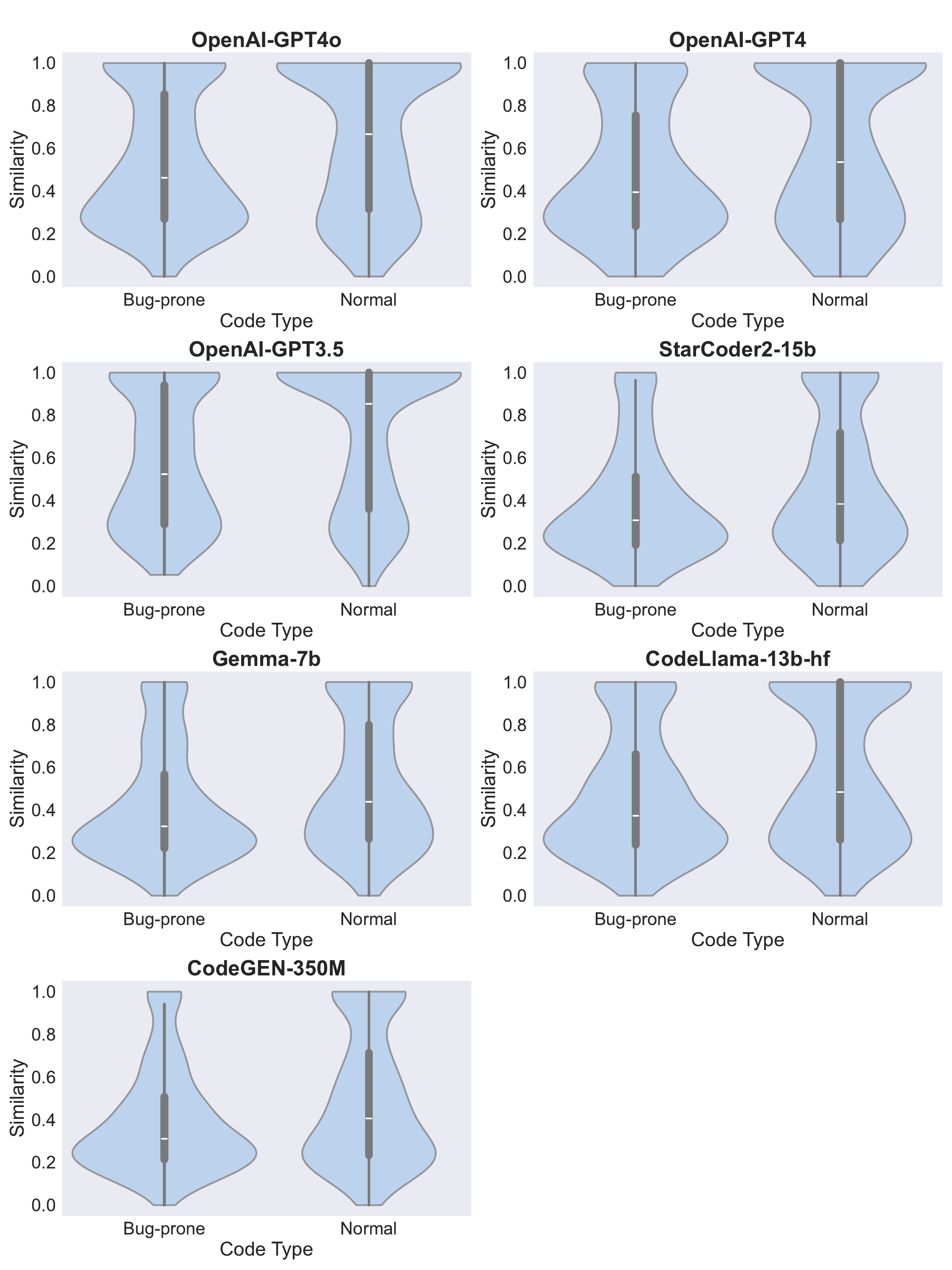}
    \caption{Distribution of similarity scores between LLM completions and ground-truth code for Bug-prone vs Normal Code code lines (Experimental Results).}
    \label{fig:similarity_distribution}
\end{figure*}

To further explore this difference in performance, we analyze the distribution of similarity scores, as shown in Fig.~\ref{fig:similarity_distribution}. The violin plots illustrate the difference in performance across various models on bug-prone and normal code tasks. For all evaluated models, including OpenAI-GPT4o, OpenAI-GPT3.5, and CodeLlama-13B-hf, normal code completions consistently yield higher similarity scores (ranging from 0.8 to 1.0) compared to bug-prone code, which typically falls between 0.2 and 0.4. This pattern is particularly evident in OpenAI-GPT4o, where normal code completions exhibit higher and more concentrated similarity scores, indicating greater reliability and accuracy in error-free contexts.

These results underscore a key limitation of current LLMs: the presence of bugs or structural complexities significantly reduces their performance on bug-prone code. This is reflected in both lower similarity scores and wider score distributions, which highlight the challenge of generating correct code in such error-prone environments.

\finding{LLMs exhibit nearly equal probabilities of generating correct and buggy code in bug-prone tasks, with a ratio of 1.01:1.00. On average, each model produces 151 correct completions and 149 buggy completions. 
Among all the bugs, the memorization rates for buggy code, where models reuse errors from their training datasets, can be as high as 82.61\%. LLMs also show significantly lower accuracy in bug-prone tasks compared to normal code (e.g., 12.27\% vs. 29.85\% for GPT-4), highlighting the increased difficulty of handling bug-prone contexts.}


\textbf{Human Evaluation.}
To further show the effectiveness of our evaluation metric, we conduct a human study to assess whether the evaluation of experienced software developers about the category (i.e., Correct-close) of single lines of code is consistent with our metric. In particular, as introduced in Sec.~\ref{Evaluation Metrics}. We do not use existing metrics like CodeBLEU because they are designed for multi-line code. In this experiment, we take CodeBERT score and CodeBLEU as baselines. We use their respective thresholds of 0.75 and 0.4, which are determined through experimental analysis. 

For the evaluation, we select buggy and fixed lines from the 546 code samples in the experiment, with each sample containing two lines (one buggy and one fixed), resulting in 1,092 lines. We exclude instances where the completion results matched the selected ones, resulting in 216 distinct code samples for evaluation, comprising a total of 432 lines. 

Participants in this experiment include graduate students majoring in computer science, information technology, and software engineering, as well as professional developers with more than three years of experience in software development, totaling 10 volunteers.
Each participant in the experiment receives a random set of 30 different code samples. Each sample includes the correct completion, the buggy one, a completion generated by an LLM, and the best completion selected after post-processing. Participants need to identify which category the completed code belongs to. We then compare their choices to the results obtained from our evaluation metrics.

The results are shown in Table~\ref{The Results Of Human Evaluation Experiment}. In this table, we find there is substantial agreement between our evaluation metrics and human judgment, confirming the reliability of our metric. The accuracy rate for Correct-close and Bug-close completions stands at 73.8\%/92.7\% (76/38 out of 103/41 samples). The overall accuracy rate is 83.7\% (251 out of 300 samples) across all categories. When compared with baselines, we find that our metric performs better than the CodeBERT score and CodeBLEU with the highest total accuracy (83.7\%). This shows the effectiveness of our metric.

\begin{table*}
    \begin{center}
    \caption{The results of our human evaluation}
    \label{The Results Of Human Evaluation Experiment}
    \resizebox{0.65\linewidth}{!}{
    \begin{tabular}{l|cccc} 
    \toprule
        & \textbf{Correct-close} & \textbf{Bug-close} & \textbf{Non-compliant} & \textbf{Total} \\
      \midrule
  \textbf{Ours} & \textbf{76/103 (73.8\%)} & 38/41 (92.7\%) & \textbf{137/156 (87.8\%)} & \textbf{251/300 (83.7\%)} \\
  CodeBERT & 65/103 (63.1\%) & \textbf{40/41 (97.6\%)} & 130/156 (83.3\%) & 235/300 (78.3\%) \\
  CodeBLEU & 73/103 (70.9\%) & 31/41 (75.6\%) & 79/156 (50.6\%) & 183/300 (61.0\%) \\
      \bottomrule
    \end{tabular}
    }
  \end{center}

\end{table*}

\finding{The human evaluation shows that our chosen metrics align well with human evaluation across a broad spectrum of code completions. The overall accuracy rate is 83.7\% across all categories. Compared with CodeBERT score and CodeBLEU, our metric still performs the best.}



\subsection{RQ2: What factors contribute to errors in bug-prone code completion by LLMs?} 
\label{sec:rq2}

To answer RQ2, we conduct an analysis focusing on the statement types of the completions. 
Our goal is to determine whether statement types influence the vulnerability of code and to pinpoint the code statements where LLMs tend to falter. Understanding the relationship between statement types and bug generation is vital. It may lead to more effective bug detection and correction strategies in automated code completion approaches, which can improve their reliability.

To achieve this goal, we use the parser generator tool tree-sitter~\cite{tree-sitter} to explore the types of these statements. We then categorized the first statement in the LLMs' completion results and their corresponding fixed code. The classification criteria rely on the root node type provided by the tree-sitter. 

\begin{table*}[t]
\centering
\caption{The code types for completion results from different models}
\label{The code types for completion results from different models}
\resizebox{\textwidth}{!}{ 
\begin{tabular}{lccccccc}
\toprule
\textbf{Statement Type} & \textbf{CodeGEN-350M} & \textbf{StarCoder2-15b} & \textbf{CodeLlama-13b-hf} & \textbf{Gemma-7b} & \textbf{OpenAI-GPT3.5} & \textbf{OpenAI-GPT4} & \textbf{OpenAI-GPT4o}\\
\midrule
Expression Statement & 15/14/59 & 12/12/54 & 19/17/49 & 16/15/64 & 21/18/46 & 17/18/48 & 8/4/14 \\
Field Declaration & 5/4/5 & 5/2/4 & 8/3/2 & 6/4/5 & 4/5/3 & 6/3/1 & 2/1/0 \\
For Statement & 1/1/8 & 0/1/10 & 1/1/11 & 1/3/12 & 1/2/10 & 2/2/12 & 0/0/2 \\
Identifier & 3/0/19 & 2/1/139 & 2/0/19 & 0/1/18 & 3/2/12 & 2/1/11 & 2/1/7 \\
If Statement & 20/13/125 & 11/7/38 & 27/22/96 & 22/16/121 & 33/23/95 & 26/18/91 & 3/3/18 \\
Variable Declaration & 12/2/41 & 11/4/38 & 23/8/50 & 14/6/47 & 29/14/45 & 16/6/65 & 4/0/11 \\
Method Invocation & 1/2/9 & 1/4/7 & 3/4/5 & 1/2/10 & 1/3/10 & 2/3/9 & 1/3/2 \\
Method Declaration & 2/6/2 & 2/3/3 & 2/6/1 & 1/4/0 & 3/4/5 & 1/8/4 & 0/2/0 \\
Return Statement & 9/16/62 & 9/14/58 & 18/19/47 & 13/14/47 & 23/19/39 & 10/24/42 & 6/8/16 \\
Throw Statement & 0/1/9 & 1/1/3 & 0/2/5 & 0/1/5 & 0/1/2 & 1/1/4 & 0/0/2 \\
While Statement & 0/0/8 & 0/0/1 & 0/1/4 & 1/1/2 & 0/0/0 & 2/0/5 & 0/0/0 \\
Other & 9/6/57 & 11/8/69 & 12/12/47 & 6/11/56 & 16/13/41 & 11/6/68 & 77/72/277 \\
\bottomrule
\end{tabular}
}

\end{table*}

Table~\ref{The code types for completion results from different models} shows the results of the completion of the code types for the line completed by LLMs. In this table, each line represents a code type and each column represents a LLM. The three numbers within the table represent Correct Completions (Correct-close + Correct-identical), buggy completions (Bug-close + Bug-identical), and Non-Compliant completions for each model. In particular, the ``other'' type includes ``Binary Expression,'' ``Boolean Type,'' ``Break Statement,'' ``Class Body,'' and other types that have fewer than five instances in each model. 

From this table, we observe that the ``if statement'' and ``expression statement'' are the most common statement types. The ``if statements'', appear frequently across all models, with a relatively high number of buggy completions, indicating a common area of difficulty. For example, OpenAI-GPT3.5 has a high number of correct (33), but also buggy (23) completions, suggesting challenges in accurately predicting conditional logic.
The expression statements show varied performance across models. Gemma-7b tends to have a higher rate of non-compliant results (64) compared to correct (16) and buggy (15), which may indicate issues with understanding or generating expressions. 

We also observe that method invocation and return statements show significant buggy and non-compliant completions across models. For ``method invocations'' and return statements, the average correct completion rates are 11.30\% and 17.77\%, illustrating the complexity and error-prone nature of ``method invocations'' and return statements in completion. 

For the proportions of correct and buggy completions, we find that in most of the types, the performance is close to their ability to generate buggy outputs, except for ``variable declarations'' and ``method invocations''. For ``variable declarations'', we observe that the models generally generate a higher proportion of correct completions compared to buggy ones (average 2.62:1.00). This suggests that models may have a better grasp of the syntactic and contextual requirements needed for accurately completing ``variable declarations''. This can be due to the relatively straightforward nature of ``variable declarations'', which often involve less complex logic than other statement types. Conversely, for ``method invocations'', the correct to buggy completion ratio is less favorable (average 1.00:2.14), indicating a challenge for the models. ``method invocations'' often involve complex dependencies, such as the correct identification of method names, understanding of the expected parameters, and the context in which the method is used. These complexities make ``method invocations'' more prone to the historical bugs, reflecting a significant challenge in the model's ability to handle intricate interactions within the code. 

\finding{For the completed code by LLMs, ``if statement'' and ``expression statement'' are the most common statement types. ``Method invocation'' and the ``return statement'' show significant buggy and non-compliant completions across models. In most of the types, the performance is close to their ability to generate buggy outputs, except for ``variable declarations'' (the proportion of correct and buggy completions is 2.62:1.00) and ``method invocations'' (1.00:2.14). }

\begin{table*}[t]
\centering
\caption{The code types of the fixed code that correspond to the completed code from different models}
\label{The code types of the fixed code that correspond to the completed code from different models}
\resizebox{\textwidth}{!}{ 
\begin{tabular}{lccccccc}
\toprule
\textbf{Code Type} & \textbf{CodeGEN-350M} & \textbf{StarCoder2-15B} & \textbf{CodeLlama-13B-hf} & \textbf{Gemma-7B} & \textbf{OpenAI-GPT3.5} & \textbf{OpenAI-GPT4} & \textbf{OpenAI-GPT4o} \\
\midrule
Expression Statement & 13/13/58 & 15/10/59 & 19/18/47 & 16/15/53 & 22/23/39 & 18/17/49 & 22/19/43 \\
Field Declaration & 5/4/5 & 5/2/7 & 9/3/2 & 6/4/4 & 4/5/5 & 6/3/5 & 5/4/5 \\
For Statement & 1/1/9 & 0/0/11 & 2/1/8 & 1/2/8 & 1/2/8 & 1/1/9 & 1/1/9 \\
Identifier & 4/0/3 & 2/1/4 & 2/0/5 & 0/2/5 & 3/2/2 & 2/2/3 & 3/2/2 \\
If Statement & 21/10/102 & 11/8/114 & 27/23/83 & 22/14/97 & 32/19/82 & 25/18/90 & 28/17/88 \\
Variable Declaration & 11/3/98 & 10/6/96 & 22/10/80 & 12/7/93 & 27/15/70 & 16/11/85 & 16/11/85 \\
Method Invocation & 1/2/9 & 1/4/7 & 3/4/5 & 1/2/9 & 1/3/8 & 2/3/7 & 1/4/7 \\
Method Declaration & 2/5/6 & 2/3/8 & 2/6/5 & 1/4/8 & 3/4/6 & 1/7/5 & 1/8/4 \\
Return Statement & 9/17/55 & 10/14/57 & 18/16/47 & 15/14/52 & 25/16/40 & 11/22/48 & 14/17/50 \\
Throw Statement & 0/1/1 & 0/1/1 & 0/1/1 & 0/0/2 & 0/1/1 & 1/0/1 & 0/1/1 \\
While Statement & 0/1/3 & 0/0/4 & 0/1/3 & 1/2/1 & 0/2/2 & 2/0/2 & 1/1/2 \\
Other & 10/8/55 & 9/8/56 & 11/12/50 & 6/12/55 & 16/12/45 & 11/6/56 & 11/9/53 \\
\bottomrule
\end{tabular}
}
\end{table*}


In Table~\ref{The code types of the fixed code that correspond to the completed code from different models}, the three numbers represent the code type for the first line of Correct Completions, buggy completions, and Non-Compliant Completions corresponding to the fixed code in the dataset. From the table, we find that in the fixed code, the ``if statement'' remains the most frequent type of code, maintaining a significant presence across all models and accounting for 31.2\% of the completions. However, there is a consistently high number of buggy and non-compliant completions for ``if statements'', which underscores a common difficulty in accurately completing conditional logic, just as in our analysis in Table~\ref{The code types for completion results from different models}.

Compared to the results in Table 4, we observe a significant difference in ``variable declarations''. In the code completion results from LLMs, these declarations occur at an average frequency of 12.6\%, while in the accurate code samples from the dataset, they make up 20.5\% of statements, ranking second only to ``if statements''. This discrepancy indicates that accurately completing ``variable declarations'' poses a challenge for LLMs, likely due to their insufficient understanding of specific programming contexts and type inference capabilities.

For other findings, we find that the results are similar to the former ones. The performance is close to their ability to generate buggy outputs.  For ``variable declarations'', the models generate a higher proportion of correct completions compared to buggy ones. For ``method invocations'', the correct-to-buggy completion ratio is still less favorable.

\finding{``if statements'' remain the most common types. The ``variable declarations'' show significant buggy and non-compliant completions across models. For the performance and ability to generate buggy outputs, the results are similar to the former.}

\subsection{RQ3: How can post-processing methods mitigate the negative effects of bug-prone code on LLM-based code completion?}

\begin{table*}[t]
\centering
\caption{The Correct Results With Post-processing Approaches}

\label{The Correct-close Results With Majority Voting Strategy}
\resizebox{\textwidth}{!}{ 
\begin{tabular}{l|ccccccc|ccc}
\toprule
\textbf{Tem.} & OpenAI-GPT4o & OpenAI-GPT4 & \textbf{OpenAI-GPT3.5} & \textbf{Gemma} & \textbf{CodeLlama} & \textbf{StarCoder2} & \textbf{CodeGEN} & \textbf{Voting} & \textbf{PE} & \textbf{Sel.} \\
\midrule 
\underline{\textbf{0.1}} & 179/363(49.3\%) & 167/338(49.4\%) & 217/402(54.0\%) & 157/311(50.5\%) & 187/366(51.1\%) & 102/231(44.2\%) & 131/286(45.8\%) & \textbf{180/369(48.8\%)} & 165/341(48.4\%)& 128/300(42.7\%)\\
0.2 & 220/395(55.7\%) & 182/364(50.0\%) & 179/338(53.0\%) & 157/308(51.0\%) & 186/351(53.0\%) & 117/237(49.4\%) & 125/284(44.0\%) & \textbf{188/359(52.4\%)} &174/346(50.3\%)& 138/307(45.0\%)\\
0.3 & 214/399(53.6\%) & 186/369(50.4\%) & 182/345(52.8\%) & 163/308(52.9\%) & 188/362(51.9\%) & 107/238(45.0\%) & 130/286(45.5\%) & \textbf{188/375(50.1\%)} &166/340(48.8\%)& 126/299(42.1\%) \\
\underline{\textbf{0.4}} & 184/365(50.4\%) & 171/336(50.9\%) & 212/383(55.4\%) & 158/306(51.6\%) & 186/353(52.7\%) & 108/228(47.4\%) & 146/287(50.9\%) & \textbf{193/371(52.0\%)} &168/337(49.9\%)& 139/297(46.8\%) \\
0.5 & 222/394(56.3\%) & 169/350(48.3\%) & 171/343(49.9\%) & 145/298(48.7\%) & 179/346(51.7\%) & 98/218(45.0\%) & 127/267(47.6\%) & \textbf{182/352(51.7\%)} &160/339(47.2\%) & 127/290(43.8\%) \\
0.6 & 206/394(52.3\%) & 195/372(52.4\%) & 179/337(53.1\%) & 146/290(50.3\%) & 178/331(53.8\%) & 102/232(44.0\%) & 121/252(48.0\%) & \textbf{185/353(52.4\%)} & 149/334(44.6\%)  & 129/287(44.9\%) \\
\underline{\textbf{0.7}} & 177/361(49.0\%) & 168/335(50.1\%) & 209/379(55.1\%) & 142/284(50.0\%) & 177/336(52.7\%) & 104/210(49.5\%) & 120/248(48.4\%) & \textbf{188/368(51.1\%)} & 160/325(49.2\%)& 142/283(50.2\%) \\
0.8 & 193/374(51.6\%) & 182/367(49.6\%) & 177/345(51.3\%) & 130/263(49.4\%) & 169/316(53.5\%) & 84/193(43.5\%) & 106/245(43.3\%) & \textbf{173/351(49.3\%)} & 141/311(45.3\%)& 123/257(47.9\%)\\
0.9 & 206/360(57.2\%) & 188/351(53.6\%) & 167/342(48.8\%) & 138/263(52.5\%) & 152/308(49.4\%) & 80/188(42.6\%) & 103/242(42.6\%) & \textbf{188/353(53.3\%)} &  155/307(50.5\%)& 109/225(48.4\%) \\
\underline{\textbf{1.0}} & 178/349(51.0\%) & 185/343(53.9\%) & 187/353(53.0\%) & 157/311(50.5\%) & 186/366(50.8\%) & 104/229(45.4\%) & 101/219(46.1\%) & \textbf{184/367(50.1\%)} & 153/335(45.7\%) &126/269(46.8\%)\\
1.1 & 169/349(48.4\%) & 172/340(50.6\%) & 177/343(51.6\%) & 123/224(54.9\%) & 141/275(51.3\%) & 91/168(54.2\%) & 85/199(42.7\%) & \textbf{176/333(52.9\%)} & 156/304(51.3\%) & 112/225(49.8\%) \\
1.2 & 165/323(51.1\%) & 182/360(50.6\%) & 173/334(51.8\%) & 95/168(56.5\%) & 119/259(45.9\%) & 54/125(43.2\%) & 79/197(40.1\%) & 149/316(47.2\%) & \textbf{135/283(47.7\%)} & 88/194(45.4\%) \\
\underline{\textbf{1.3}} & 179/349(51.3\%) & 187/343(54.5\%) & 159/301(52.8\%) & 65/152(42.8\%) & 115/206(55.8\%) & 41/85(48.2\%) & 87/189(46.0\%) & 149/294(50.7\%) & \textbf{150/294(51.0\%)} &87/183(47.5\%) \\
1.4 & 155/277(56.0\%) & 160/330(48.5\%) & 171/317(53.9\%) & 51/95(53.7\%) & 83/167(49.7\%) & 23/57(40.4\%) & 63/140(45.0\%) & 123/257(47.9\%) & \textbf{129/249(51.8\%)}& 69/138(50.0\%) \\
1.5 & 118/218(54.1\%) & 161/327(49.2\%) & 151/305(49.5\%) & 28/63(44.4\%) & 80/144(55.6\%) & 18/40(45.0\%) & 62/125(49.6\%) & 114/229(49.8\%) & 122/234(52.1\%) & \textbf{69/119(58.0\%)} \\
\underline{\textbf{1.6}} & 152/297(51.2\%) & 156/287(54.4\%) & 98/178(55.1\%) & 20/44(45.5\%) & 62/110(56.4\%) & 6/12(50.0\%) & 48/106(45.3\%) & \textbf{103/183(56.3\%)} & 94/193(48.7\%)&53/105(50.5\%) \\
\midrule
\textbf{Voting} & \textbf{185/368(50.3\%)} & 177/345(51.3\%) & \textbf{220/403(54.6\%)} & \textbf{165/327(50.5\%)} & 192/373(51.5\%) & \textbf{126/267(47.2\%)} & \textbf{151/299(50.5\%)}  & && \\
\textbf{PE} & 169/350(48.2\%) & \textbf{171/329(52.0\%)} & 185/365(50.7\%) & 159/318(50.0\%) & 174/330(52.7\%) & 118/268(44.0\%) & 125/264(47.3\%) & &&\\
\textbf{Sel.} & 145/316(45.9\%) & 165/320(51.6\%) & 162/300(54.0\%) & 97/197(49.2\%) & \textbf{124/229(54.1\%)} & 73/166(44.0\%) & 86/183(47.0\%) & && \\

\bottomrule
\end{tabular} 
}

\end{table*}

To answer RQ3, we review and analyze the methods we employ. Our objective is to select the optimal completion from multiple code completion results through post-processing approaches. To this end, We conduct a set of experiments, which involved selecting results from different parameters under the same model and from the same parameters across different models. As introduced in Sec.~\ref{Post-processing}, we consider three different post-processing approaches, namely the voting mechanism (Voting), the prompt engineering (PE), and the selection model (Sel.). For each technique, we focus on two aspects: (1) selecting the best completion from multiple code completion results generated by different models, and (2) selecting the best completion from multiple code completion results generated at different temperatures.



\paragraph{Models} We first conduct experiments on voting among different models, where we aim to select the most similar results from different models with the same temperature setting across the previous seven models. 

Table~\ref{The Correct-close Results With Majority Voting Strategy} provides detailed results. In this table, each line represents a temperature and each column represents a LLM. The three numbers within the table represent Correct Completions (Correct-close + Correct-identical), Correct and buggy completions (Correct-close + Correct-identical + Bug-close + Bug-identical), and their proportions. The last three columns represent the performance of three post-processing approaches on all code completion results generated by different models in a specific temperature.

Based on the results in Table~\ref{The Correct-close Results With Majority Voting Strategy}, we find that the Voting strategy generally achieves the best results among the three post-processing approaches. It consistently secures the highest average number of correct completions (166, compared to 149 for PE and 114 for the Sel.) and the highest average rate of correct completions (51.0\%, compared to 48.9\% for PE and 47.5\% for the Sel.). This indicates that Voting is particularly effective in pooling insights from various models to enhance the overall accuracy of code completions.

When examining peak performance, the Sel. stands out by achieving the highest rate of correct completions at 58.0\% at a temperature setting of 1.5. When optimally tuned to specific conditions, the Sel. can outperform other strategies in maximizing completion accuracy. 

However, when considering the original models, except the Sel. on temperature 1.5, all other post-processing approaches fail to outperform the best-performing model in either the number of correctness or the proportions. 
This observation highlights a limitation of the post-processing approaches: while it does reduce the risk of extremely poor outputs and ensures a more consistent baseline performance, it may not attain the maximum potential accuracy that a single, optimally functioning model could achieve under specific conditions.

We also find that the performance of post-processing approaches is still close to their ability to generate buggy outputs. This means that it still has half of the completions on the bug-prone code, which may contain faults. Therefore, despite the implementation of post-processing approaches like Voting, PE, and Sel., there remains a significant challenge in fully mitigating the tendency of LLMs to generate incorrect outputs. This persistence of error generation, particularly in bug-prone code scenarios, can be problematic, since even a single mistake, when propagated through an automated system, can result in substantial bugs in the final software product.

\begin{table*}[t]
\centering
\caption{The Results on Two-lines Completion With Post-processing Approaches}
\label{dis:twolinemodel}
\resizebox{\textwidth}{!}{ 
\begin{tabular}{l|ccccccc|ccc}
\toprule
\textbf{Tem.} & \textbf{OpenAI-GPT4o} & \textbf{OpenAI-GPT4} & \textbf{OpenAI-GPT3.5} & \textbf{Gemma-7B} & \textbf{CodeLlama-13B-hf} & \textbf{StarCoder2-15B} & \textbf{CodeGEN-350M} & \textbf{Voting} & \textbf{PE} & \textbf{Sel.}\\
\midrule
\underline{\textbf{0.1}} & 129/253(51.0\%) & 179/358(50.0\%) & 227/417(54.4\%) & 183/336(54.5\%) & 202/386(52.3\%) & 116/246(47.2\%) & 148/318(46.5\%) & \textbf{182/352(51.7\%)} & 111/239(46.4\%) & 71/186(38.2\%) \\
0.2 & 133/268(49.6\%) & 187/351(53.3\%) & 235/414(56.8\%) & 173/330(52.4\%) & 209/382(54.7\%) & 123/249(49.4\%) & 140/317(44.2\%) & \textbf{181/347(52.2\%)} & 123/241(51.0\%) & 95/191(49.7\%) \\
0.3 & 137/258(53.1\%) & 194/367(52.9\%) & 229/412(55.6\%) & 189/347(54.5\%) & 192/378(50.8\%) & 129/253(51.0\%) & 150/320(46.9\%) & \textbf{190/359(52.9\%)} & 115/239(48.1\%) & 90/193(46.6\%) \\
\underline{\textbf{0.4}} & 120/257(46.7\%) & 180/352(51.1\%) & 221/401(55.1\%) & 178/336(53.0\%) & 200/380(52.6\%) & 121/246(49.2\%) & 155/310(50.0\%) & \textbf{192/358(53.6\%)} & 122/247(49.4\%) & 91/190(47.9\%) \\
0.5 & 133/251(53.0\%) & 185/356(52.0\%) & 236/409(57.7\%) & 167/318(52.5\%) & 197/377(52.3\%) & 112/234(47.9\%) & 139/299(46.5\%) & \textbf{184/342(53.8\%)} & 99/228(43.4\%) & 81/184(44.0\%) \\
0.6 & 135/269(50.2\%) & 191/363(52.6\%) & 209/395(52.9\%) & 168/318(52.8\%) & 186/351(53.0\%) & 113/238(47.5\%) & 140/294(47.6\%) & \textbf{188/343(54.8\%)} & 101/236(42.8\%) & 86/194(44.3\%) \\
\underline{\textbf{0.7}} & 135/273(49.5\%) & 183/358(51.1\%) & 211/394(53.6\%) & 173/316(54.7\%) & 199/368(54.1\%) & 109/224(48.7\%) & 146/295(49.5\%) & \textbf{186/356(52.2\%)} & 111/239(46.4\%) & 84/183(45.9\%) \\
0.8 & 144/276(52.2\%) & 182/362(50.3\%) & 199/392(50.8\%) & 146/297(49.2\%) & 187/346(54.0\%) & 89/207(43.0\%) & 122/267(45.7\%) & \textbf{168/329(51.1\%)} & 98/220(44.5\%) & 83/173(48.0\%) \\
0.9 & 140/272(51.5\%) & 180/360(50.0\%) & 209/376(55.6\%) & 165/285(57.9\%) & 167/334(50.0\%) & 82/193(42.5\%) & 121/265(45.7\%) & \textbf{174/340(51.2\%)} & 108/221(48.9\%) & 73/166(44.0\%) \\
\underline{\textbf{1.0}} & 130/266(48.9\%) & 202/362(55.8\%) & 193/380(50.8\%) & 180/334(53.9\%) & 202/387(52.2\%) & 119/239(49.8\%) & 114/255(44.7\%) & \textbf{181/349(51.9\%)} & 102/239(42.7\%) & 99/206(48.1\%) \\
1.1 & 127/272(46.7\%) & 190/359(52.9\%) & 193/362(53.3\%) & 128/234(54.7\%) & 147/279(52.7\%) & 98/175(56.0\%) & 98/225(43.6\%) & 161/318(50.6\%) & \textbf{112/215(52.1\%)} & 83/180(46.1\%) \\
1.2 & 134/269(49.8\%) & 184/344(53.5\%) & 184/346(53.2\%) & 109/184(59.2\%) & 146/276(52.9\%) & 59/133(44.4\%) & 105/223(47.1\%) & \textbf{146/298(49.0\%)} & 85/199(42.7\%) & 59/161(36.6\%) \\
\underline{\textbf{1.3}} & 130/259(50.2\%) & 191/344(55.5\%) & 153/291(52.6\%) & 77/153(50.3\%) & 123/221(55.7\%) & 47/89(52.8\%) & 99/199(49.7\%) & \textbf{153/283(54.1\%)} & 96/202(47.5\%) & 59/145(40.7\%) \\
1.4 & 123/247(49.8\%) & 178/319(55.8\%) & 160/279(57.3\%) & 54/98(55.1\%) & 89/178(50.0\%) & 25/49(51.0\%) & 82/162(50.6\%) & 121/242(50.0\%) & \textbf{84/164(51.2\%)} & 46/106(43.4\%) \\
1.5 & 123/244(50.4\%) & 158/316(50.0\%) & 131/218(60.1\%) & 24/49(49.0\%) & 81/145(55.9\%) & 15/29(51.7\%) & 67/143(46.9\%) & \textbf{106/204(52.0\%)} & 84/168(50.0\%) & 50/103(48.5\%) \\
\underline{\textbf{1.6}} & 107/240(44.6\%) & 152/286(53.1\%) & 99/182(54.4\%) & 16/29(55.2\%) & 59/107(55.1\%) & 5/8(62.5\%) & 66/133(49.6\%) & \textbf{81/145(55.9\%)} & 55/132(41.7\%) & 36/86(41.9\%) \\
\midrule

\textbf{Voting} & \textbf{150/281(53.4\%)} & \textbf{101/191(52.9\%)} & 82/166(49.4\%) & 102/213(47.9\%) & 94/202(46.5\%) & 75/172(43.6\%) & \textbf{110/212(51.9\%)} & &&\\
\textbf{PE} & 170/351(48.4\%) & 173/333(52.0\%) & 195/373(52.3\%) & \textbf{155/302(51.3\%)} & \textbf{179/329(54.4\%)} & \textbf{125/264(47.3\%)} & 130/264(49.2\%) & &&\\
\textbf{Sel.} & 160/335(47.8\%) & 168/326(51.5\%) & \textbf{162/300(54.0\%)} & 108/231(46.8\%) & 119/237(50.2\%) & 61/137(44.5\%) & 90/188(47.9\%) & &&\\

\bottomrule
\end{tabular}
}

\end{table*}

\paragraph{Different Temperatures}

We then focus on selecting the best completion from multiple code completion results generated by each model in different temperatures.

The results are also in Table~\ref{The Correct-close Results With Majority Voting Strategy}, where the last three rows represent the performance of the models at different temperatures (0.1, 0.4, 0.7, 1.0, 1.3, 1.6). We chose these 7 temperatures to keep the selection manageable and avoid overwhelming complexity.

From the results, we find that the Voting strategy typically yields the best overall results among the three techniques, with the highest average number of correct completions (173) and the highest average correct rate (51\%). For optimal performance, PE is prominent, particularly when applied exclusively with OpenAI-GPT3.5, achieving the best correct rate of 57.5\%. This high performance is likely due to the model's ability to synergize well with tailored prompts that enhance its inherent strengths, thus optimizing the quality of code completions.

Although the post-processing techniques reach a peak effectiveness of 57.5\% in generating correct completions, their overall utility remains constrained by a strong correlation with the production of incorrect outputs. This persistent issue underscores the inherent challenges in mitigating errors through post-processing alone.

\finding{Although Majority Voting achieves the best overall results among the three post-processing approaches, the implementation of these existing techniques does not fundamentally alter the fact that the performance of code completion strategies is still close to their propensity to generate incorrect outputs. }

\section{Extended Analysis and Discussion}
\label{sec:discussion}


In this section, we extend our analysis and present a discussion of our findings through several key aspects. First, we explore the impact of hyper-parameter tuning on model performance, particularly focusing on temperature and token length settings. Second, we extend our analysis to recent code bugs using the ConDefects dataset to validate our findings' generalizability. Third, we investigate the performance of LLMs on multi-line code completion tasks to examine whether our findings hold beyond single-line completions. Additionally, we analyze emerging reasoning LLMs to assess their behavior in bug-prone contexts. Through these discussions, we aim to provide deeper insights into the challenges and opportunities in bug-prone code completion using LLMs.

\subsection{Hyper-Parameter Tuning}

We notice that the hyper-parameters in the LLMs are also important and may influence the performance of them. In this section, we explore the influence of the models with different hyper-parameters. 

\begin{figure}[h]
\centering
\includegraphics[width=0.5\textwidth]{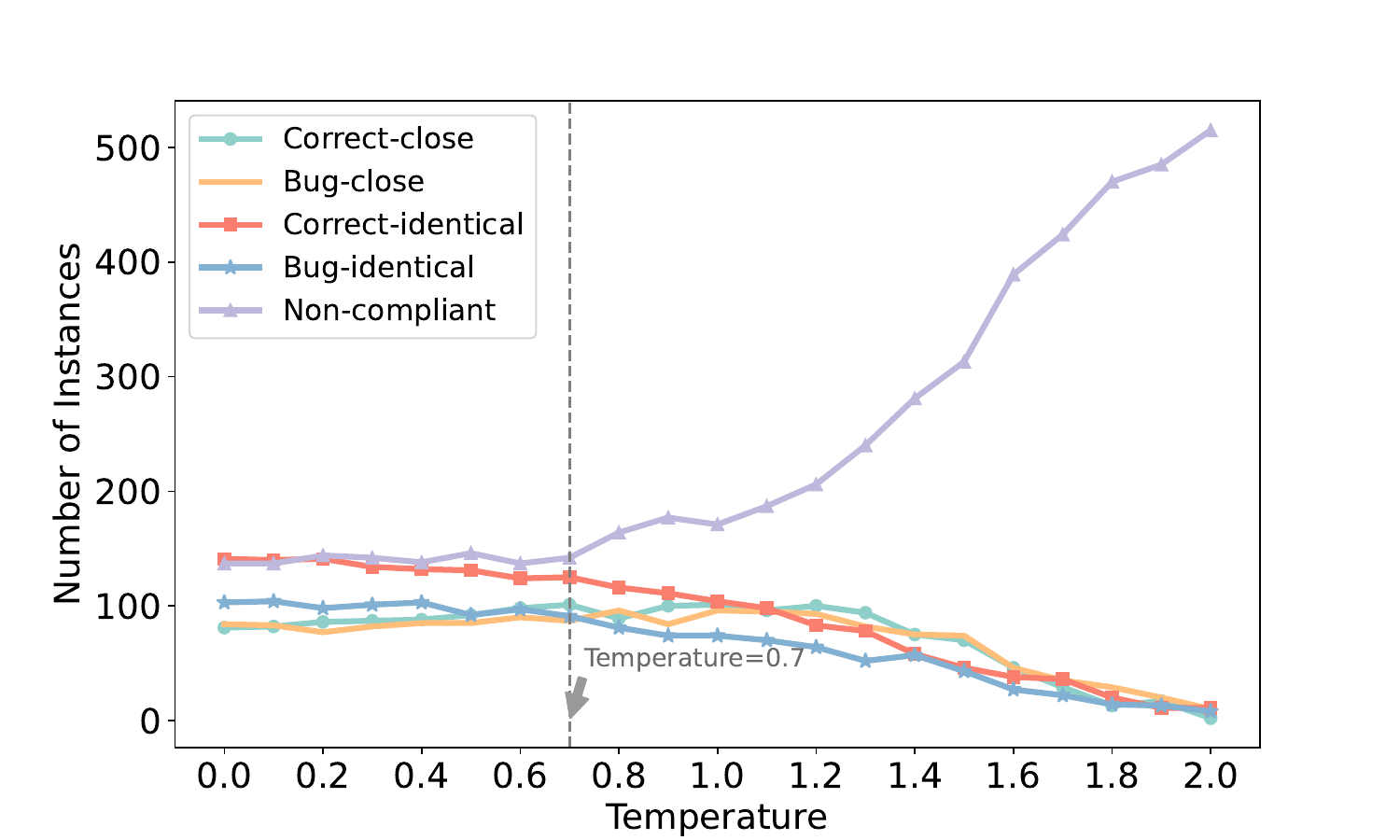}
\caption{Exploring the Impact of Temperature Settings on OpenAI-GPT3.5 Code Completion}
\label{temperature}
\end{figure}

\begin{figure*}[ht]
\centering
\includegraphics[width=0.75\textwidth]{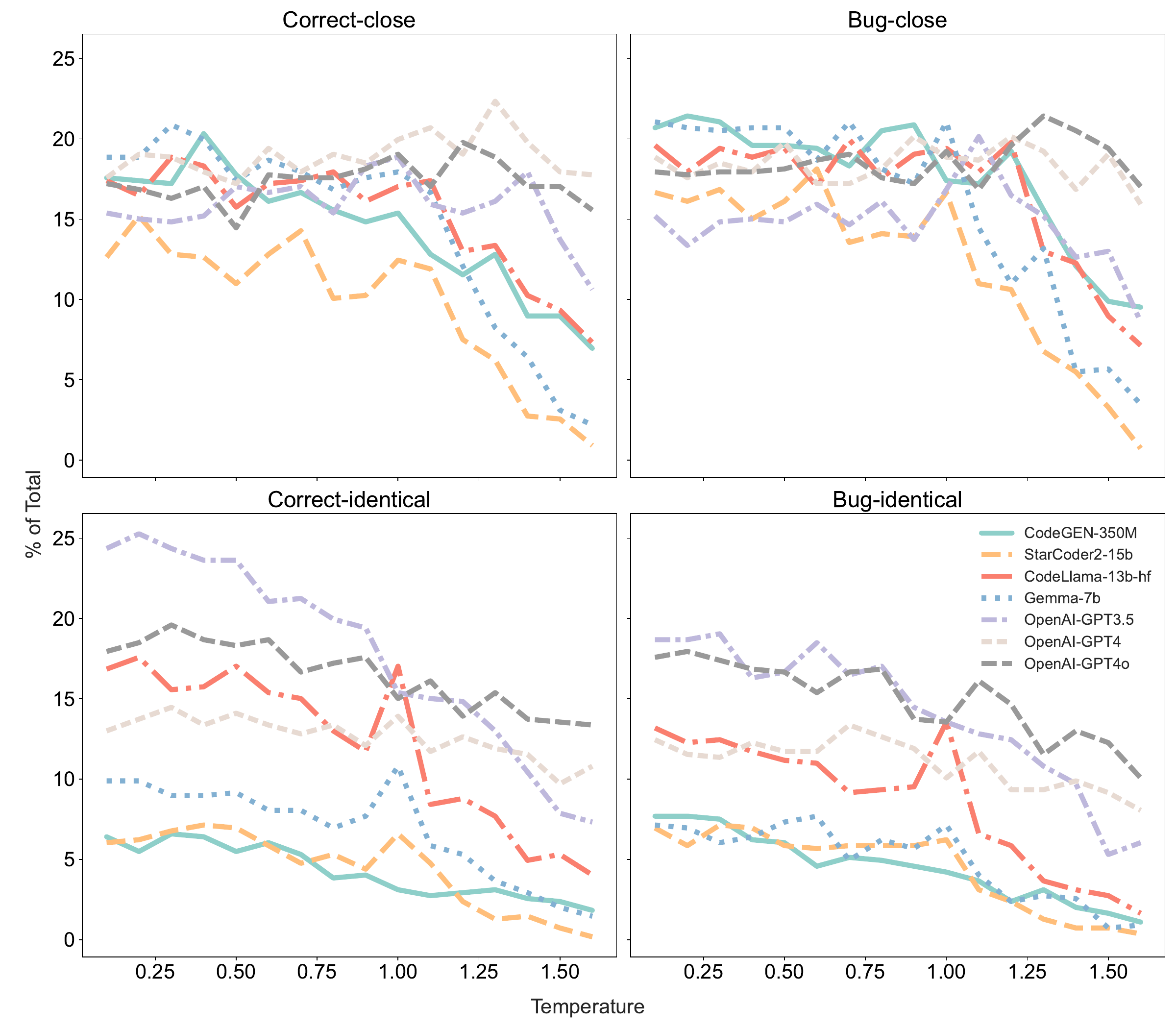}
\caption{Exploring the Impact of Temperature Settings on Code Completion for Various Large Language Models}
\label{temperature_all_model}
\end{figure*}

\paragraph{Temperature} For the temperature, we conduct additional experiments on all seven previously mentioned models. By altering the temperature parameter, we observe each model's code completion performance at various temperature settings. We show detailed results on OpenAI-GPT3.5 in Fig.~\ref{temperature}. We chose OpenAI-GPT3.5 for our preliminary experiments because of its popularity and proven effectiveness in code completion tasks. The results are shown in Fig.~\ref{temperature_all_model}.

From Fig.~\ref{temperature}, we find that as the temperature increases, the overall number of Non-compliant instances rises, whereas correct and buggy completions decline.  This indicates that increasing temperatures induce a greater degree of variability and a reduction in precision within code completions, alongside a diminished tendency to replicate learned patterns from the training data. 
The optimal code completion quality is achieved with the temperature setting at approximately 0.7, marking the peak in Correct-identical completions. The rate of Correct-close completions is notably high at this setting, constituting 41.3\% of the total. Conversely, the proportion of Bug-close code was only 32.6\%, indicating that at this temperature, the model completes code of relatively higher quality, offering more Correct-close and fewer Bug-close completions. This may be attributed to the model's enhanced ability to balance accuracy and creativity at this temperature, thereby achieving better code completion results. The quality of code completions began to decline when the temperature exceeded 0.7. This downward trend is caused by greater divergence and randomness in outputs as temperature increases, leading to more completions that were neither accurately correct nor buggy. 
The trends observed in Fig.~\ref{temperature_all_model} persist across the models, although the specific temperature at which peak performance is achieved varies.

In Fig.~\ref{temperature} and Fig.~\ref{temperature_all_model}, we also find that despite the temperature changes, the previous findings remain consistent: 1) the well-performing models in generating correct code are also prone to generating substantial incorrect outputs; 2) their performance is closely mirrored by their ability to generate buggy outputs; and 3) existing LLMs tend to generate a large amount of off-target, irrelevant, or completely nonsensical completions. For the finding of code memorization, we get a new observation that as the temperature increases, existing models display a decreased propensity to regenerate learned patterns from their training datasets.

\begin{figure}
\centering
\includegraphics[width=0.5\textwidth]{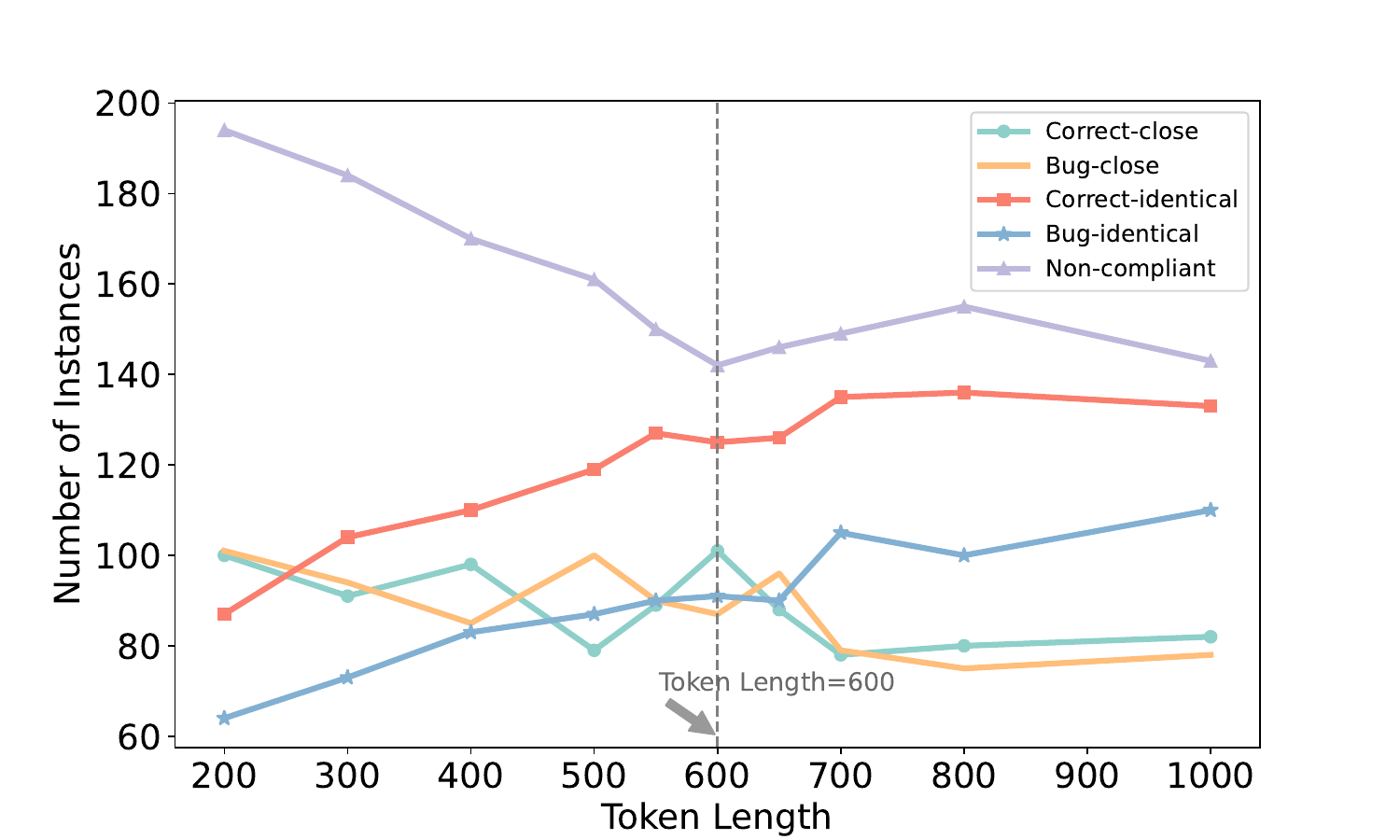}
\caption{Exploring the Impact of Different Token Lengths on Code Completion}
\label{token_len}
\end{figure}

\paragraph{Token Length} 
For the token length, we conduct additional experiments on OpenAI-GPT3.5, varying parameters including temperature and token length. Token length denotes the context's length given to the model, with a longer length implying more information in the provided request. The experimental results are shown in Fig.~\ref{token_len}. 

The results show that irrespective of token length, there is a consistent pattern of model results aligning with buggy codes, and this misalignment becomes more obvious as the token length increases. Specifically, at a token length of 600, the model finds a sweet spot, achieving an optimal balance between code completion quality and economic efficiency, with the best results achieved using fewer tokens. 
For the previous findings (Finding 1 to Finding 4), the results are still consistent.


\subsection{Extension to Recent Code Bugs}
To further validate our findings and explore their generalizability to more recent code issues, we conduct additional experiments using the ConDefects dataset~\cite{wu2024condefects}, which contains coding faults from October 2021 to September 2023. This dataset is particularly interesting as it helps mitigate potential data leakage risks by including only recent faults that are less likely to appear in LLMs' training data.

Our results on ConDefects, as shown in Table~\ref{ConDefects completion results}, demonstrate similar patterns to those observed with Defects4J, reinforcing our earlier findings:

\begin{itemize}
    \item The correct-to-incorrect completion ratio remains problematic. For example, OpenAI-GPT4o shows 28.54\% correct completions (11.16\% Correct-identical + 17.38\% Correct-close) versus 29.86\% buggy completions (6.16\% Bug-identical + 23.70\% Bug-close), maintaining a nearly 1:1 ratio.
    \item Models continue to struggle with complex code structures, as evidenced by the high proportion of non-compliant outputs across all models (ranging from 24.54\% to 61.00\%).
    \item Even the best-performing models show significant error rates. For instance, Gemma-7B, despite having a relatively high number of Correct-identical completions (145), still produces 577 buggy or non-compliant outputs.
\end{itemize}

These consistent findings across both Defects4J and ConDefects datasets suggest that the challenges we identified in bug code completion are not specific to older code patterns but represent fundamental limitations in current LLM approaches. The persistence of these patterns, even with more recent code examples, reinforces the need for improved methods specifically designed to handle bug-prone code contexts, regardless of the code's age or origin.

\begin{table*}[t]
\centering
\caption{Analysis of code quality completed by different LLMs on ConDefects dataset}

\resizebox{.8\linewidth}{!}{ 
\begin{tabular}{lccccc}
\toprule
\textbf{LLMs} & 
\textbf{\begin{tabular}[c]{@{}c@{}}Completely\\ Correct\end{tabular}} & \textbf{\begin{tabular}[c]{@{}c@{}}Completely\\ Faulty\end{tabular}} &
\textbf{\begin{tabular}[c]{@{}c@{}}Nearly\\ Correct\end{tabular}} & \textbf{\begin{tabular}[c]{@{}c@{}}Nearly\\ Faulty\end{tabular}} &  
\textbf{Non-compliant} \\ \midrule
OpenAI-GPT4o & 212(11.16\%) & 117(6.16\%) & 330(17.38\%) & 450(23.70\%) & 790(41.60\%) \\ 
Gemma-7B & 145(7.64\%) & 112(5.90\%) & 262(13.80\%) & 465(24.49\%) & 915(48.18\%) \\ 

\bottomrule
\end{tabular}
}
\label{ConDefects completion results}
\end{table*}

\subsection{Multi-lines Completion} In the experiments, we focus on completing the immediate next line on the bug-prone code. In practical applications
of code completion, especially in the context of integrated development environments or other software development tools, developers often focus primarily on the immediate next line suggested by the code
completion feature. 
In this section, we extend the immediate next-line completions to two-line code completions. We conduct the same experiments as in Section~\ref{sec:results}.

\begin{table*}
\centering
\caption{The Results on Two-lines Completion With Post-processing Approaches (Different Models)}
\label{dis:twolinemodel}
\resizebox{\textwidth}{!}{ 
\begin{tabular}{l|ccccc|ccc}
\toprule
\textbf{Tem.} & \textbf{OpenAI-GPT3.5} & \textbf{Gemma-7B} & \textbf{CodeLlama-13B-hf} & \textbf{StarCoder2-15B} & \textbf{CodeGEN-350M} & \textbf{Voting} & \textbf{PE} & \textbf{Sel.}\\
\midrule
0.1 & 227/417(54.4\%) & 183/336(54.5\%) & 202/386(52.3\%) & 116/246(47.2\%) & 148/318(46.5\%) & 182/352(51.7\%) & 111/239(46.4\%) & 71/186(38.2\%) \\
0.2 & 235/414(56.8\%) & 173/330(52.4\%) & 209/382(54.7\%) & 123/249(49.4\%) & 140/317(44.2\%) & 181/347(52.2\%) & 123/241(51.0\%) & 95/191(49.7\%) \\
0.3 & 229/412(55.6\%) & 189/347(54.5\%) & 192/378(50.8\%) & 129/253(51.0\%) & 150/320(46.9\%) & 190/359(52.9\%) & 115/239(48.1\%) & 90/193(46.6\%) \\
0.4 & 221/401(55.1\%) & 178/336(53.0\%) & 200/380(52.6\%) & 121/246(49.2\%) & 155/310(50.0\%) & 192/358(53.6\%) & 122/247(49.4\%) & 91/190(47.9\%) \\
0.5 & 236/409(57.7\%) & 167/318(52.5\%) & 197/377(52.3\%) & 112/234(47.9\%) & 139/299(46.5\%) & 184/342(53.8\%) & 99/228(43.4\%) & 81/184(44.0\%) \\
0.6 & 209/395(52.9\%) & 168/318(52.8\%) & 186/351(53.0\%) & 113/238(47.5\%) & 140/294(47.6\%) & 188/343(54.8\%) & 101/236(42.8\%) & 86/194(44.3\%) \\
0.7 & 211/394(53.6\%) & 173/316(54.7\%) & 199/368(54.1\%) & 109/224(48.7\%) & 146/295(49.5\%) & 186/356(52.2\%) & 111/239(46.4\%) & 84/183(45.9\%) \\
0.8 & 199/392(50.8\%) & 146/297(49.2\%) & 187/346(54.0\%) & 89/207(43.0\%) & 122/267(45.7\%) & 168/329(51.1\%) & 98/220(44.5\%) & 83/173(48.0\%) \\
0.9 & 209/376(55.6\%) & 165/285(57.9\%) & 167/334(50.0\%) & 82/193(42.5\%) & 121/265(45.7\%) & 174/340(51.2\%) & 108/221(48.9\%) & 73/166(44.0\%) \\
1.0 & 193/380(50.8\%) & 180/334(53.9\%) & 202/387(52.2\%) & 119/239(49.8\%) & 114/255(44.7\%) & 181/349(51.9\%) & 102/239(42.7\%) & 99/206(48.1\%) \\
1.1 & 193/362(53.3\%) & 128/234(54.7\%) & 147/279(52.7\%) & 98/175(56.0\%) & 98/225(43.6\%) & 161/318(50.6\%) & 112/215(52.1\%) & 83/180(46.1\%) \\
1.2 & 184/346(53.2\%) & 109/184(59.2\%) & 146/276(52.9\%) & 59/133(44.4\%) & 105/223(47.1\%) & 146/298(49.0\%) & 85/199(42.7\%) & 59/161(36.6\%) \\
1.3 & 153/291(52.6\%) & 77/153(50.3\%) & 123/221(55.7\%) & 47/89(52.8\%) & 99/199(49.7\%) & 153/283(54.1\%) & 96/202(47.5\%) & 59/145(40.7\%) \\
1.4 & 160/279(57.3\%) & 54/98(55.1\%) & 89/178(50.0\%) & 25/49(51.0\%) & 82/162(50.6\%) & 121/242(50.0\%) & 84/164(51.2\%) & 46/106(43.4\%) \\
1.5 & 131/218(60.1\%) & 24/49(49.0\%) & 81/145(55.9\%) & 15/29(51.7\%) & 67/143(46.9\%) & 106/204(52.0\%) & 84/168(50.0\%) & 50/103(48.5\%) \\
1.6 & 99/182(54.4\%) & 16/29(55.2\%) & 59/107(55.1\%) & 5/8(62.5\%) & 66/133(49.6\%) & 81/145(55.9\%) & 55/132(41.7\%) & 36/86(41.9\%) \\
\midrule
Avg. & 193/354(54.6\%)	& 133/248(53.7\%)	& 162/306(53\%)	& 85/176(49.7\%)	& 118/252(47.2\%)	& 162/310(52.3\%)	& 100/214(46.8\%)	& 74/165(44.6\%) \\

\bottomrule
\end{tabular}
}
\end{table*}

\begin{table*}
\centering
\caption{The Results on Two-lines Completion With Post-processing Approaches (Different Temperature)}
\label{dis:twolinetemp}
\resizebox{\textwidth}{!}{ 
\begin{tabular}{l|ccccc|c}
\toprule
\textbf{Tem.} & \textbf{OpenAI-GPT3.5} & \textbf{Gemma-7B} & \textbf{CodeLlama-13B-hf} & \textbf{StarCoder2-15B} & \textbf{CodeGEN-350M} & \textbf{Avg.}\\
\midrule
0.1 & 122/235(51.9\%) & 142/276(51.4\%) & 136/275(49.5\%) & 93/202(46.0\%) & 130/275(47.3\%) & 125/253(49.2\%)\\
0.4 & 118/236(50.0\%) & 141/282(50.0\%) & 138/279(49.5\%) & 99/200(49.5\%) & 138/275(50.2\%) & 127/254(49.8\%)\\
0.7 & 119/242(49.2\%) & 139/267(52.1\%) & 146/282(51.8\%) & 95/191(49.7\%) & 130/264(49.2\%) & 126/249(50.4\%)\\
1.0 & 126/261(48.3\%) & 133/267(49.8\%) & 138/274(50.4\%) & 96/195(49.2\%) & 104/232(44.8\%) & 119/246(48.5\%)\\
1.3 & 104/205(50.7\%) & 64/135(47.4\%) & 99/187(52.9\%) & 43/80(53.8\%) & 93/183(50.8\%) & 81/158(51.1\%)\\
1.6 & 77/144(53.5\%) & 9/21(42.9\%) & 49/93(52.7\%) & 5/8(62.5\%) & 58/122(47.5\%) & 40/78(51.8\%)\\
\midrule
Voting & 82/166(49.4\%) & 102/213(47.9\%) & 94/202(46.5\%) & 75/172(43.6\%) & 110/212(51.9\%) & 93/193(47.9\%)\\
PE & 195/373(52.3\%) & 155/302(51.3\%) & 179/329(54.4\%) & 125/264(47.3\%) & 130/264(49.2\%) & 157/306(50.9\%)\\
Sel. & 162/300(54.0\%) & 108/231(46.8\%) & 119/237(50.2\%) & 61/137(44.5\%) & 90/188(47.9\%) & 108/219(48.7\%)\\
\bottomrule
\end{tabular}
}
\end{table*}

The experimental results are shown in Table~\ref{dis:twolinemodel} and Table~\ref{dis:twolinetemp}. These two tables represent the results of different models on different temperatures and the results selected using the majority voting (Voting),
prompt engineering (PE), and selection model (Sel.) from code completions of various models
under specific temperatures. In this table, each line represents a temperature and each column
represents a LLM. The three numbers within the table represent Correct Completions (Nearly
Correct + Correct-identical), Correct and buggy completions (Correct-close + Completely
Correct + Bug-close + Bug-identical), and its proportions. In particular, the post-processing approaches in Table~\ref{dis:twolinemodel}
select the most similar results from different models with the same temperature setting across the
previous five models, whereas those in Table~\ref{dis:twolinetemp} select the best completion from multiple code completion results generated by each model in different temperatures.

As shown, extending the scope of code completion from a single immediate line to two lines yields consistent results with previous findings. Even well-performing models continue to tend to generate substantial incorrect outputs. The performance of code completion strategies remains closely linked to their propensity to generate errors. Additionally, the application of post-processing approaches does not significantly alter these results. The consistency of these results across different settings highlights the persistent challenges in enhancing the accuracy and reliability of automated code completions.

\subsection{Reasoning LLM Analysis} 
\label{sec:reasoning}


New reasoning models (e.g., ChatGPT-o1~\cite{reason-llms}, DeepSeek-R1~\cite{guo2025deepseek}) have attracted significant media attention due to their promising capabilities in code completion tasks. However, because code completion scenarios demand extremely high speed, these models are typically excluded from early experimental phases owing to their computational demands. To investigate whether the code completions generated by these new reasoning models behave similarly to those produced by established LLMs in bug-prone contexts, we conduct additional experiments using DeepSeek-R1.

\begin{table*}[h]
\centering
\caption{Analysis of code quality completed by DeepSeek R1}

\resizebox{0.9\linewidth}{!}{ 
\begin{tabular}{llrrrrrrr}
\toprule
\textbf{LLMs} & \textbf{Task Type} &
\textbf{\begin{tabular}[c]{@{}c@{}}Completely\\ Correct\end{tabular}} & \textbf{\begin{tabular}[c]{@{}c@{}}Completely\\ Buggy\end{tabular}} &
\textbf{\begin{tabular}[c]{@{}c@{}}Nearly\\ Correct\end{tabular}} & \textbf{\begin{tabular}[c]{@{}c@{}}Nearly\\ Buggy\end{tabular}} &  
\textbf{Non-compliant} &  
\textbf{\begin{tabular}[c]{@{}c@{}}Completely Buggy\\ Ratio\end{tabular}} \\ \midrule
DeepSeek R1 & bug-prone & 76 (13.92\%) & 65 (11.90\%) & 116 (21.25\%) & 113 (20.70\%) & 176 (32.23\%) & 36.52\% \\ 
 & normal & 163 (29.85\%) & - & 186 (34.07\%) & - & 197 (36.08\%) & - \\
\bottomrule
\end{tabular}
}
\label{Inference Model Analysis completion results}
\end{table*}


The experimental results are presented in \textbf{Table~\ref{Inference Model Analysis completion results}}, which details the code quality analysis of DeepSeek-R1 in bug-prone code completion tasks. The table is consistent with that in RQ1, which reinforces the trends observed across different LLMs. 

From this table, we observe that the key findings still hold for reasoning models. Specifically, DeepSeek-R1 exhibits a nearly balanced distribution of correct and buggy completions in bug-prone tasks. It produces 76 completely correct completions (13.92\%) and 65 completely buggy completions (11.90\%), along with 116 nearly correct (21.25\%) and 113 nearly buggy completions (20.70\%). Notably, the completely buggy ratio stands at 36.52\%, indicating a substantial error rate when handling bug-prone code. When completing normal code, DeepSeek-R1 achieves higher correctness with 163 completely correct completions (29.85\%) and 186 nearly correct completions (34.07\%). These results show that even state-of-the-art reasoning models still struggle with bug-prone code completion.

\subsection{Implications}

The findings from this study have important implications for both the development and application of LLMs in software engineering in the realm of code completion:

\paragraph{Enhanced Understanding of LLM Capabilities and Limitations} This research deepens our understanding of LLMs' capabilities and limitations in handling bug-prone code. Despite their proficiency in generating syntactically correct code, LLMs often do not fully grasp the semantic nuances, leading to potential errors. This dual nature underscores the need for enhanced model training that goes beyond syntactic understanding to include more robust semantic analysis.

\paragraph{Need for Improved Error Handling} This study underscores the need for improved error handling in code completions, especially in fragile contexts. The high incidence of errors calls for the development of more sophisticated error detection and correction mechanisms within LLMs. There is a clear need for models that not only generate code but also evaluate and adapt their outputs based on potential logical and runtime errors.

\paragraph{Refinement of Post-processing Approaches} This investigation highlights the need for refining post-processing approaches in LLM outputs. While techniques like majority voting, prompt engineering, and selective modeling have been employed to refine the outputs of LLMs, their effectiveness remains limited. The findings suggest that these techniques, although useful in increasing the reliability of completions, still fall short of significantly reducing errors. This highlights an opportunity for developing more advanced post-processing algorithms that can better discern and correct inaccuracies in model outputs.

\paragraph{Practical Integration in Development Environments} This research advocates for the practical integration of enhanced LLMs with robust post-processing tools in integrated development environments (IDEs). For practitioners, such integration can significantly boost productivity and reduce error rates in real-time coding scenarios. However, the integration process must be handled with care to avoid introducing additional complexity or overhead that can detract from the user experience.


\section{Threats to Validity}
\label{sec:threats}

\textbf{Threats to external validity} lie in LLM code completion capabilities, dataset quality, and result generalization. We selecte top-performing and popular LLMs, including seven models specifically for this study, alongside others such as Bloom et al.\cite{le2023bloom}, Vicuna et al.\cite{zheng2024judging}, ChatGLM ~\cite{du2022glm, zeng2022glm} and Phi et al.\cite{li2023textbooks}. For data sets, we mainly used the widely recognized Defects4J benchmark\cite{sarhan_survey_2022, chen2023future, durieux2019empirical, 7985681}, and further validated our findings using the ConDefects dataset in our discussion section. To prevent data leakage in training data, we excluded projects from both datasets in the bug-fix dataset~\cite{zhu2021syntax}. Although our experiments are confined to Java code, the comprehensive and real-world nature of both datasets supports their reliability and generalization for evaluating LLM performance variations. 

\noindent
\textbf{Threats to internal validity } lie in code completion, which involves selecting model parameters and comparing code lengths. In our process, we use default settings for all parameters except the temperature setting. While varying default parameters might induce result variations, limiting adjustments to the temperature parameter ensured experimental consistency and control, mitigating factors that may impact internal validity. For evaluation simplicity and to emphasize initial accuracy, we compare only the first line of generated code. As code completion generally operates on a token-by-token and line-by-line basis, the first line provides a reliable benchmark for assessing the accuracy and relevance of the completion tool.

\noindent
\textbf{Threats to construct validity } lie in the choice of evaluation metrics. Metrics like CodeBLEU and CodeBERT score, designed for larger code blocks, often falter in precision when applied to single lines. To enhance accuracy, we use a metric that merges the LCS with LED. We validated this metric through a human evaluation, which confirmed its alignment with human judgment, thus showing its effectiveness.

\section{Conclusion}
\label{sec:conclusion}

In this study, we conduct the first  investigation into the ability of large language models (LLMs) to handle bug-prone code completion tasks. Our empirical evaluation on seven state-of-the-art LLMs, including ChatGPT (3.5, 4.0, 4o), CodeLlama, StarCoder, CodeGEN, and Gemma, reveals several critical insights. We find that completing bug-prone code is significantly more difficult than standard code completion, with models generating correct and buggy completions with nearly equal probability (approximately 151 correct versus 149 buggy completions on average) and achieving substantially lower accuracy (e.g., 12.27\% versus 29.85\% for GPT-4). A striking discovery was that 44.44\% of the bugs produced by LLMs are identical to historical bugs, indicating a tendency to replicate past errors rather than learn from corrected versions. Additionally, certain code constructs, such as method invocations, return statements, and variable declarations, prove particularly error-prone, while existing post-processing techniques fail to significantly reduce these error rates, despite improving output consistency. These findings highlight LLM limitations in managing complex code dependencies, underscoring the need for research. High error rates and bug replication suggest new methodologies are essential.

\bibliographystyle{IEEEtran}
\bibliography{main}

\end{document}